\documentclass[twoside,11pt]{article}

\usepackage[accepted,specialissue]{melba}

\usepackage{microtype}
\usepackage[utf8]{inputenc}
\usepackage[T1]{fontenc}
\usepackage[english]{babel}
\usepackage{dsfont}
\usepackage{bm}
\usepackage{amsfonts}
\usepackage{ulem}
\usepackage{etoolbox}
\usepackage{nicefrac}
\usepackage{pifont}
\usepackage{mathtools}
\usepackage{chemarrow}
\usepackage{booktabs}
\usepackage{adjustbox}
\usepackage{multirow}
\usepackage{makecell}
\usepackage[ruled]{algorithm2e}
\usepackage{enumitem}
\usepackage{hyperref}
\usepackage{afterpage}
\usepackage{subcaption}
\usepackage{longtable}
\usepackage{url}
\usepackage[capitalise]{cleveref}
\usepackage[toc,title,page]{appendix}

\DeclareMathOperator{\x}{\mathbf{x}}
\DeclareMathOperator{\z}{\mathbf{z}}
\DeclareMathOperator{\E}{\mathbb{E}}
\DeclareMathOperator{\R}{\mathbb{R}}
\DeclareMathOperator{\eps}{\bm{\epsilon}}

\melbaid{2023:016}  
\doi{https://doi.org/10.59275/j.melba.2023-fbe4}
\melbaauthors{Fu et al.}  
\volume{2}
\firstpageno{507}  
\melbayear{2023}  
\datesubmitted{10/2023}  
\datepublished{12/2023}  

\melbaspecialissue{Special Issue for Generative Models}
\melbaspecialissueeditors{Mert Sabuncu, Sotirios A. Tsaftaris}

\ShortHeadings{Recycling for Medical Image Segmentation with Diffusion Denoising Models}{Fu et al.}

\title{A Recycling Training Strategy for Medical Image Segmentation with Diffusion Denoising Models}

\author{\firstname Yunguan \surname Fu \orcid{0000-0002-1184-7421} \email yunguan.fu.18@ucl.ac.uk; y.fu@instadeep.com \\  
	\addr University College London, UK; InstaDeep, UK
	\AND
	\firstname Yiwen \surname Li \orcid{0000-0002-7794-9391} \email yiwen.li@st-annes.ox.ac.uk \\
	\addr University of Oxford, UK
        \AND
	\firstname Shaheer U. \surname Saeed \orcid{0000-0002-5004-0663} \email shaheer.saeed.17@ucl.ac.uk \\
	\addr University College London, UK
        \AND
	\firstname Matthew J. \surname Clarkson \orcid{0000-0002-5565-1252} \email m.clarkson@ucl.ac.uk \\
	\addr University College London, UK
        \AND
	\firstname Yipeng \surname Hu \orcid{0000-0003-4902-0486} \email yipeng.hu@ucl.ac.uk \\
	\addr University College London, UK
}

\begin{document}

\maketitle

\begin{abstract}
Denoising diffusion models have found applications in image segmentation by generating segmented masks conditioned on images. Existing studies predominantly focus on adjusting model architecture or improving inference, such as test-time sampling strategies. In this work, we focus on improving the training strategy and propose a novel recycling method. During each training step, a segmentation mask is first predicted given an image and a random noise. This predicted mask, which replaces the conventional ground truth mask, is used for the denoising task during training. This approach can be interpreted as aligning the training strategy with inference by eliminating the dependence on ground truth masks for generating noisy samples. Our proposed method significantly outperforms standard diffusion training, self-conditioning, and existing recycling strategies across multiple medical imaging data sets: muscle ultrasound, abdominal CT, prostate MR, and brain MR. This holds for two widely adopted sampling strategies: denoising diffusion probabilistic model and denoising diffusion implicit model. Importantly, existing diffusion models often display a declining or unstable performance during inference, whereas our novel recycling consistently enhances or maintains performance. We show for the first time that, under a fair comparison with the same network architectures and computing budget, the proposed recycling-based diffusion models achieved on-par performance with non-diffusion-based supervised training. Furthermore, by ensembling the proposed diffusion model and the non-diffusion counterpart, significant improvements to the non-diffusion models have been observed across all applications, demonstrating the value of this novel training method. This paper summarizes these quantitative results and discusses their values, with a fully reproducible JAX-based implementation, released at \url{https://github.com/mathpluscode/ImgX-DiffSeg}.
\end{abstract}


\section{Introduction}

Diffusion denoising models, first proposed by~\citet{sohl2015deep,ho2020denoising,ho2022classifier}, are generative models that produce data samples through iterative denoising processes. These models achieved superior performance compared to generative adversarial networks~\citep{goodfellow2020generative} and became the foundation for many image generation applications such as DALL$\cdot$E 2~\citep{ramesh2022hierarchical}, stable diffusion, and Midjourney~\citep{rombach2022high}, etc. Given the success in computer vision, diffusion models have been adapted in medical imaging in various fields, including image synthesis~\citep{dorjsembe2022three,khader2022medical}, image denoising~\citep{hu2022unsupervised}, anomaly detection~\citep{wolleb2022diffusion_anomaly}, classification~\citep{yang2023diffmic}, segmentation~\citep{wu2022medsegdiff,rahman2023ambiguous}, and registration~\citep{kim2022diffusemorph}. Among these, segmentation is one of the most foundational tasks in medical imaging and a variety of applications have been explored, including liver CT~\citep{xing2023diff}, lung CT~\citep{zbinden2023stochastic,rahman2023ambiguous}, abdominal CT~\citep{wu2023medsegdiff,fu2023importance}, brain MR~\citep{pinaya2022fast,wolleb2022diffusion_ensemble,wu2023medsegdiff,xing2023diff,bieder2023diffusion}, and prostate MR~\citep{fu2023importance}.

For segmentation tasks, although various model architectures and training strategies ~\citep{wang2022medical} have been proposed, U-net equipped with attention mechanisms and trained by supervised learning consistently remains the state-of-the-art model and an important baseline. In comparison, divergent observations have emerged: some studies reported superior performance of diffusion-based segmentation models~\citep{amit2021segdiff,wu2022medsegdiff,wu2023medsegdiff,xing2023diff}, while others observed the opposite trend~\citep{pinaya2022fast,wolleb2022diffusion_ensemble,kolbeinsson2022multi,fu2023importance}. This inconsistency may result from different training schemes, network architectures, and application-specific modifications in comparisons, suggesting that challenges persist in applying diffusion models for image segmentation.

Formally, conditioning on an image, diffusion-based segmentation models operate by progressive denoising, starting with random noise and ultimately yielding the corresponding segmentation masks. In comparison to their non-diffusion counterparts, the necessity of supplementary noisy masks as input leads to increased memory demands that can pose challenges, particularly for processing 3D volumetric medical images. To address this, volume slicing~\citep{wu2023medsegdiff} or patching~\citep{xing2023diff,bieder2023diffusion} has been used to manage memory limitations. However, diffusion model training still requires considerable computation due to its inherent iterative nature, since the same model needs to learn to denoise masks with varying levels of noise. Consequently, enhancing the diffusion model performance while adhering to a fixed compute budget is of significant importance. Empirically, using the reparametrisation~\citep{kingma2021variational}, the denoising training task has shifted from noise prediction~\citep{wolleb2022diffusion_ensemble,wu2022medsegdiff} to mask prediction~\citep{fu2023importance,zbinden2023stochastic} due to the superior performance and faster learning. Furthermore,~\citet{fu2023importance} highlighted a limitation of diffusion models, noting the misalignment between training and inference procedures, since training samples were generated from ground truth masks. This raises concerns of data leakage as discussed in~\citet{chen2022generalist}. However, there have been limited studies in medical image segmentation that rigorously compare diffusion models with their non-diffusion counterparts and examine diffusion training efficiency.

In this work, we present a substantial extension to the preliminary work~\citep{fu2023importance} and focus on an improvement in the diffusion denoising model training strategy that applies to 2D and 3D medical image segmentation in different modalities. First, a novel recycling approach has been introduced. Different from~\citet{fu2023importance}, in the first step during training, the input is completely corrupted by noise instead of a partially corrupted ground truth. This seemingly minor adjustment effectively eliminates the ground truth information from model inputs, which further aligns the training strategy toward inference. The proposed diffusion models can refine or maintain segmentation accuracy throughout the inference process. On the contrary, all other diffusion models demonstrate declining or unstable performance trends. Our research showcases the superior performance of our method compared to established diffusion training strategies~\citep{ho2020denoising,chen2022analog,watson2023novo,fu2023importance} for both denoising diffusion probabilistic model-based~\citep{ho2020denoising} and denoising diffusion implicit model-based~\citep{song2020denoising} sampling procedures. We also achieved on-par performance with non-diffusion baselines that had not been observed in the previous study~\citep{fu2023importance}. Second, we introduce an ensemble model that averages the predicted probabilities from the proposed diffusion-based model and non-diffusion counterpart, resulting in significant improvement to the non-diffusion baseline. Third, we extended the experiments to four large data sets -- 2D muscle ultrasound with $3910$ images, 3D abdominal CT with $300$ images, 3D prostate MR with $589$ images, and 3D brain MR with $1251$ images, further demonstrating the robustness of the proposed method against different applications and data types. Lastly, we integrated a Transformer block into our network architecture. This brings our models in line with contemporary state-of-the-art approaches, rendering our findings more pertinent to real-world applications. To mitigate the increased memory consumption resulting from this addition, we employed patch-based training and inference strategies. The JAX-based framework has been released on \url{https://github.com/mathpluscode/ImgX-DiffSeg}.

\section{Related Works}

The diffusion process is a Markov process where data structures are gradually noise-corrupted and eventually destroyed (noising process). A reverse diffusion process (denoising process) can then be learned, where the objective is to gradually recover the data structure.~\citet{sohl2015deep} first proposed diffusion models which map the disrupted data to a noise distribution.~\citet{ho2020denoising} have shown that such modeling is equivalent to score-matching models, a class of models that estimates the gradient of the log-density~\citep{hyvarinen2005estimation,vincent2011connection,song2019generative,song2020improved}. This led to a simplified variational lower bound training objective and a denoising diffusion probabilistic model (DDPM)~\citep{ho2020denoising}. DDPM achieved state-of-the-art performance for unconditional image generation on CIFAR10 at the time. In practice, DDPMs were found suboptimal on log-likelihood estimation and~\citet{nichol2021improved} addressed this with a learnable variance schedule, sinusoidal noise schedule, and an importance sampling for time steps. Furthermore, diffusion models were trained with hundreds or thousands of steps, inference with the same number of steps is time-consuming. Therefore, different strategies have been proposed to enable faster sampling. While~\citet{nichol2021improved} suggested variance resampling without modifying the probabilistic distribution,~\citet{song2020denoising} derived a deterministic model, denoising diffusion implicit model (DDIM), which shares the same marginal distribution as DDPM.~\citet{liu2022pseudo} further generalized the reverse step of DDIM into an ordinary differential equation and used high-order numerical methods (e.g., Runge-Kutta method) with predicted noise to perform sampling with second-order convergence. Besides,~\citet{zheng2022truncated,lyu2022accelerating,guo2022accelerating} accelerated diffusion model training by shortening the noising schedule and only considering a truncated diffusion chain with less noise. These unconditioned denoising diffusion models have been successfully applied in multiple medical imaging applications~\citep{kazerouni2023diffusion}, including brain MR image generation~\citep{dorjsembe2022three,khader2022medical}, optical coherence tomography denoising~\citep{hu2022unsupervised}, and chest X-ray pleural effusion detection~\citep{wolleb2022diffusion_anomaly}.

Guided diffusion models have been developed to generate data in a controllable manner.~\citet{song2020score,dhariwal2021diffusion} used gradients of pre-trained classifiers to bias the sampling process, without modifying the diffusion model training.~\citet{ho2022classifier}, on the other hand, modified the models to take additional information as input, enabling end-to-end conditional diffusion model training. For medical image synthesis, conditions can be patient biometric information~\citep{pinaya2022brain}, genotypes data~\citep{moghadam2023morphology}, or images from different modalities~\citep{saeed2023bi}. Conditional diffusion models have also been explored for medical image classification~\citep{yang2023diffmic}, segmentation~\citep{wu2022medsegdiff,rahman2023ambiguous}, and registration~\citep{kim2022diffusemorph}. Particularly for image segmentation, the diffusion models apply the noising and denoising on the segmentation masks, and the network takes a noisy mask and an image to perform denoising.

In contrast to the continuous spectrum of values found in natural images, image segmentation mask values are categorical and nominal. The Gaussian-based continuous diffusion processes behind DDPM and DDIM cannot be directly applied.~\citet{chen2022analog} therefore encoded categories with binary bits and relaxed them to real values for continuous diffusion models.~\citet{han2022ssd,fu2023importance} encoded categories with one-hot embeddings and performed diffusion on scaled values.~\citet{li2022diffusion,strudel2022self} encoded the discrete data and applied diffusion processes in embedding spaces directly. Alternatively, discrete diffusion models have been proposed to model the transition matrix between categories based on discrete probability distributions, including binomial distribution~\citep{sohl2015deep}, categorical distribution~\citep{hoogeboom2021argmax,austin2021structured,gu2022vector}, and Bernoulli distribution~\citep{chen2023berdiff}. In this work, we follow~\citet{fu2023importance} to perform diffusion on scaled binary masks.

Originally, DDPM models were trained through noise prediction~\citep{ho2020denoising}, where the loss was calculated between the predicted and sampled noises. Many diffusion-based segmentation models directly adopted this strategy~\citep{wolleb2022diffusion_ensemble,wu2022medsegdiff}. Alternatively,~\citet{kingma2021variational} derived an equivalent formulation (often known as $\x_0$ reparameterization) of the variational lower bound and simplified the loss to a norm between predicted data and the corresponding ground truth. For segmentation, this is equivalent to predicting the segmentation mask for each time step. Compared to noise prediction, multiple studies found that this mask prediction strategy is more efficient~\citep{fu2023importance,wang2023dformer,lai2023denoising}. Furthermore,~\citet{chen2022analog} suggested self-conditioning to use these predictions as additional input to improve diffusion models for image synthesis. Self-conditioning contains two steps: the first step predicts a noise-free sample given a noise-corrupted sample only; the second step uses the same timestep and inputs the same noise-corrupted sample, as well as the prediction from the first step. This technique was later adopted for protein design~\citep{watson2023novo} with an additional reverse step, where the second step performs denoising in a smaller timestep where the noise level is lower. However, in both cases, the noisy samples are directly derived from the ground truth, which is not available during inference. This risks data leakage during training and empirically leads to overfitting and lack of generalization as discussed in~\citet{chen2022generalist,kolbeinsson2022multi,lai2023denoising}.~\citet{chen2022generalist,young2022sud} addressed this issue by controlling the signal-to-noise ratio so that less information is preserved after noising:~\citet{chen2022generalist} scaled the mask value ranges to implicitly amplify the noise level, and~\citet{young2022sud} explicitly varied the scale and standard deviation of the Gaussian noise added to the masks. On the other hand,~\citet{kolbeinsson2022multi} proposed recursive denoising that iterates through each step during training, without using ground truth as input. However, such a strategy extends the training length by a factor of hundreds or more, making it practically infeasible for larger 3D medical image data sets. Following these studies,~\citet{fu2023importance} concluded that the lack of generalization in diffusion-based segmentation models is due to the misalignment between training and inference processes.~\citet{fu2023importance} thus presented a two-step recycling training strategy: the first step ingests a partially noisied sample for mask prediction; the predicted mask is then noise-corrupted again for denoising training. Compared to recursive denoising, this method requires a limited training time increase. This method also resembles PD-DDPM~\citep{guo2022accelerating}, where a pre-segmentation is used for noising. However, PD-DDPM requires a separate pre-segmentation network and more device memory, thus not suitable for 3D image segmentation applications. 

\section{Background}

\subsection{Denoising Diffusion Probabilistic Model}
\begin{align}\label{eq:noising-denoising}
    \x_T \autorightleftharpoons{}{} \cdots \autorightleftharpoons{}{} \x_t \autorightleftharpoons{$p_\theta(\x_{t-1}\mid\x_t)$}{$q(\x_t\mid\x_{t-1})$} \x_{t-1} \autorightleftharpoons{}{} \cdots \autorightleftharpoons{}{} \x_0
\end{align}
\paragraph{Definition} The denoising diffusion probabilistic models (DDPM)~\citep{ho2020denoising} consider a \textit{forward} process (illustrated from right to left in~\Cref{eq:noising-denoising}): given a sample $\mathbf{x}_0\sim q(\mathbf{x}_0)$, a noise-corrupted sample $\mathbf{x}_t$ follows a multivariate normal distribution at timestep $t\in\{1,\cdots,T\}$, $q(\mathbf{x}_t\mid\mathbf{x}_{t-1})=\mathcal{N}(\mathbf{x}_t;\sqrt{1-\beta_t}\mathbf{x}_{t-1}, \beta_t\mathbf{I})$, where $\beta_t\in[0,1]$. As Gaussians are closed under convolution, given $\mathbf{x}_0$, $\mathbf{x}_t$ can be directly sampled from $\mathbf{x}_0$ as follows, $q(\x_t\mid\x_0) = \mathcal{N}(\x_t;\sqrt{\bar{\alpha}_t}\x_0, (1-\bar{\alpha}_t)\mathbf{I})$. Correspondingly, a \textit{reverse} process  (illustrated from left to right in~\Cref{eq:noising-denoising}) denoises $\mathbf{x}_{t}$ at each step, for $t\in\{T,\cdots,1\}$, $p_\theta(\mathbf{x}_{t-1}\mid\mathbf{x}_{t})=\mathcal{N}(\mathbf{x}_{t-1};\bm{\mu}_\theta(\mathbf{x}_t,t),\sigma_t^2\mathbf{I})$, with a predicted mean $\bm{\mu}_\theta(\mathbf{x}_t,t)$ and variance $\sigma_t^2\mathbf{I}$. $\sigma_t$ is a pre-defined schedule dependent on timestep $t$.  In this work, $\sigma_t^2=\tilde{\beta}_t=\frac{1 - \bar{\alpha}_{t-1}}{1-\bar{\alpha}_{t}}\beta_t$ with $\alpha_t = 1-\beta_t$ and $\bar{\alpha}_t = \prod_{s=1}^t \alpha_s$. The mean $\bm{\mu}_\theta(\mathbf{x}_{t},t)=\frac{\sqrt{\bar{\alpha}_{t-1}}\beta_{t}}{1-\bar{\alpha}_{t}}\hat{\x}_0+\frac{1 - \bar{\alpha}_{t-1}}{1-\bar{\alpha}_{t}}\sqrt{\alpha_{t}}\mathbf{x}_{t}$, also know as $\x_0$ parameterization, where $\hat{\x}_0$ is an estimation of $\x_0$ from a learned neural network $\hat{\x}_0=f_\theta(\mathbf{x}_t,t)$.

\paragraph{Training}
For each step during training, a noise-corrupted sample $\x_t$ is sampled and input to the neural network $f_\theta$ to predict $\x_0$. The network is then trained with loss $L_\text{denoising}(\theta)=\E_{t,\x_0,\x_t}L(\x_{0},\hat{\x}_0)$ with $t$ sampled from $1$ to $T$. $L(\cdot,\cdot)$ is a loss function in the space of $\x$. In this work, importance sampling~\citep{nichol2021improved} is used for time step $t$, where the weight for $t$ is proportional to $\E_{\x_0,\x_t}L(\x_{0},\hat{\x}_0)$.
\begin{subequations}\label{eq:ddpm-train}
\begin{align}
  \x_t &\sim \mathcal{N}(\x_t;\sqrt{\bar{\alpha}_t}\mathbf{x}_0, (1-\bar{\alpha}_t)\mathbf{I}),&(\text{Sampling})\\
    \hat{\mathbf{x}}_0 &= f_\theta(t,\mathbf{x}_t),&(\text{Prediction})\\
    L_\text{denoising}(\theta)
    &=\E_{t,\x_0,\x_t}L(\x_{0},\hat{\mathbf{x}}_0),&(\text{loss calculation})
\end{align}
\end{subequations}

\paragraph{Inference} At inference time, the denoising starts with a randomly sampled Gaussian noise $\x_T\sim\mathcal{N}(\bm{0},\bm{I})$ and the data is denoised step-by-step for $t=T,\cdots,1$:
\begin{align*}
    p_\theta(\x_{t-1}\mid\x_{t})&= \mathcal{N}(\x_{t-1};\bm{\mu}_\theta(\x_t,t),\sigma_t^2\mathbf{I})\\
    \bm{\mu}_\theta(\x_t,t&)=\frac{\sqrt{\bar{\alpha}_{t-1}}\beta_t}{1-\bar{\alpha}_{t}}\hat{\x}_0
    +\frac{1 - \bar{\alpha}_{t-1}}{1-\bar{\alpha}_{t}}\sqrt{\alpha_t}\x_{t}
\end{align*}
Optionally, the variance schedule $\beta_t$ can be down-sampled to reduce the number of inference steps~\citep{nichol2021improved}. A detailed review of DDPM and the loss has been summarised in~\Cref{app:ddpm} and we refer the readers to~\citet{sohl2015deep,ho2020denoising,nichol2021improved,kingma2021variational} and other literature for in-depth understanding and derivations.
\subsection{Diffusion for Segmentation}
When applying diffusion models for segmentation, noising and denoising are performed on the segmentation masks. The ground-truth binary mask, where channels correspond to classes that include the background, is denoted by $\mathbf{x}_0$. For the $i$-th pixel/voxel, the value for the $j$-th channel is in $1$ if it belongs to class $j$ and $-1$ otherwise. The training process (illustrated in~\Cref{fig:method_comparison}) is similar to~\Cref{eq:ddpm-train} except that the segmentation network $f_\theta(I,\mathbf{x}_t,t)$ now takes the image $I$ as an additional input for prediction $\hat{\x}_0$. $L(\cdot,\cdot)$ is a weighted sum of cross entropy and foreground-only Dice loss.
\begin{subequations}\label{eq:ddpm-train-segmentation}
\begin{align}
  \x_t &\sim \mathcal{N}(\x_t;\sqrt{\bar{\alpha}_t}\mathbf{x}_0, (1-\bar{\alpha}_t)\mathbf{I}),&(\text{Sampling})\\
    \hat{\mathbf{x}}_0 &= f_\theta(I, t,\mathbf{x}_t),&(\text{Prediction})\\
    L_\text{denoising}(\theta)
    &=\E_{t,\x_0,\x_t}L(\x_{0},\hat{\mathbf{x}}_0),&(\text{loss calculation})
\end{align}
\end{subequations}

\section{Methods}
\afterpage{
\begin{figure}[!ht]
\centering
\includegraphics[width=\linewidth]{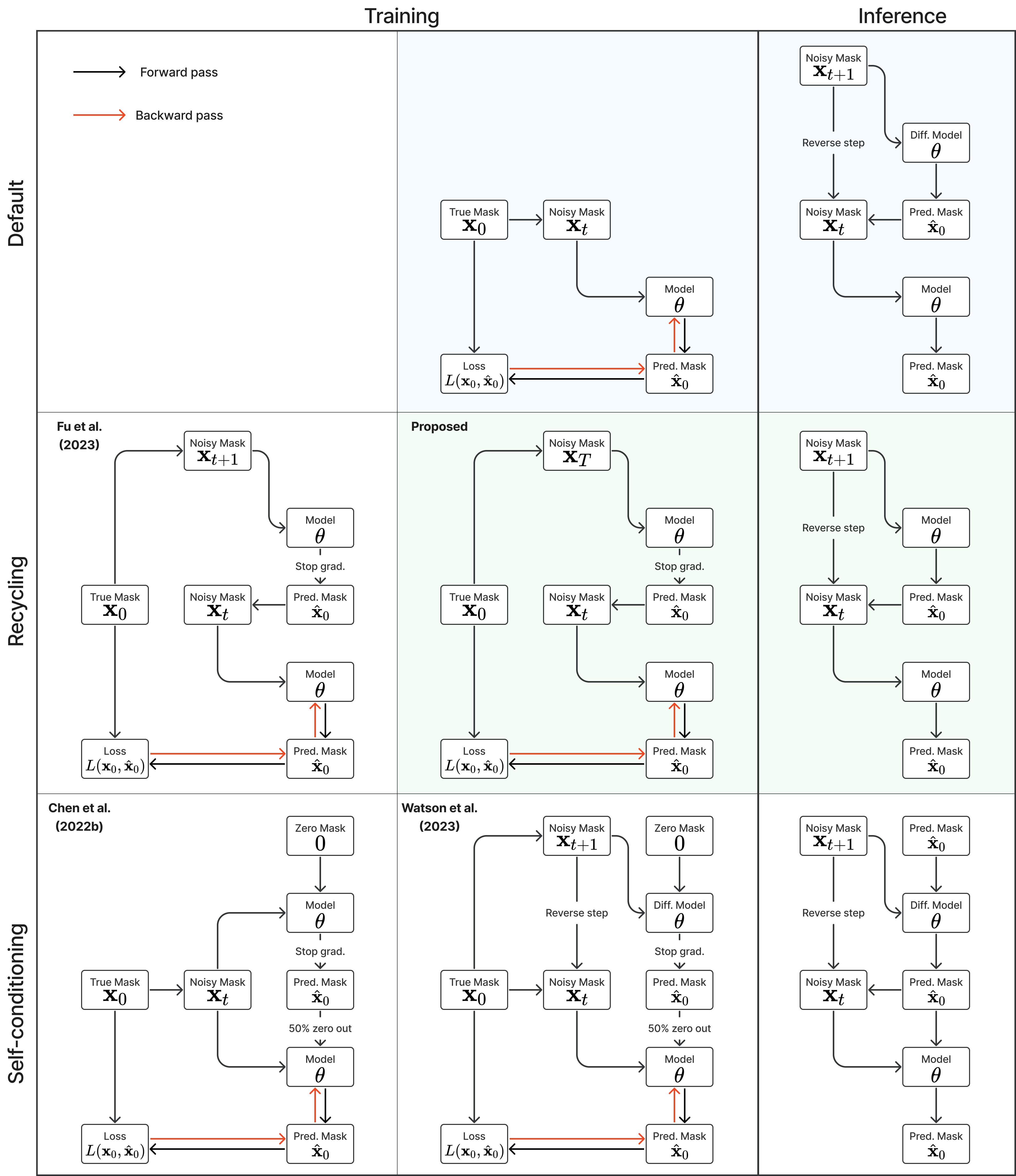}
\caption{\textbf{Illustration of training and inference processes.} The top, middle, and bottom rows show the training and inference steps for default diffusion (highlighted in blue), diffusion with recycling, and diffusion with self-conditioning, respectively. For training, different settings are presented for recycling and self-conditioning. The proposed method is highlighted in green. Notably, recycling shares the same inference steps as default diffusion, while self-conditioning is different as a result of the additional input. ``Pred.'' and ``Diff.'' stands for predicted and diffusion, respectively.} \label{fig:method_comparison}
\end{figure}
\clearpage
}
At each training step, the recycling considers a sampled time step $t<T$ and a data sample $\x_0$. First, a noise-corrupted sample $\x_T$ at time step $T$ is sampled, with $\sqrt{\bar{\alpha}_T}\approx0$. $\x_T$ is fed to the network $f_\theta$ to perform a prediction $\hat{\mathbf{x}}_0=f_\theta(I,T,\mathbf{x}_T)$. This prediction is then noise-corrupted to generate $\x_t$. A second prediction $\hat{\mathbf{x}}_0=f_\theta(I,t,\mathbf{x}_t)$ (overriding the $\hat{\mathbf{x}}_0$ for simplicity) is produced and used for loss calculation (see~\Cref{fig:method_comparison}). Formally, at each timestep $t$, the proposed recycling (denoted as ``Diff. rec. $\x_T$'') has the following steps.
\begin{subequations}\label{eq:ddpm-train-segmentation-recycling-max}
\begin{align}
    \x_T &\sim \mathcal{N}(\x_T;\sqrt{\bar{\alpha}_T}\x_0, (1-\bar{\alpha}_T)\mathbf{I}),&(\text{rec.}~\x_T,~\text{step 1, sampling}) \label{eq:rec-max-step1-sample}\\
    \hat{\mathbf{x}}_0 &= \text{StopGradient}(f_\theta(I,T,\mathbf{x}_T)),&(\text{rec.}~\x_T,~\text{step 1, prediction}) \label{eq:rec-max-step1-pred}\\
    \x_t &\sim \mathcal{N}(\x_t;\sqrt{\bar{\alpha}_t}\hat{\mathbf{x}}_0, (1-\bar{\alpha}_t)\mathbf{I}),&(\text{rec.}~\x_T,~\text{step 2, sampling}) \label{eq:rec-max-step2-sample}\\
    \hat{\mathbf{x}}_0 &= f_\theta(I,t,\mathbf{x}_t),&(\text{rec.}~\x_T,~\text{step 2, prediction}) \label{eq:rec-max-step2-pred}\\
    L_\text{denoising}(\theta)
    &=\E_{t,\x_0,\x_t}L(\x_{0},\hat{\mathbf{x}}_0),&(\text{loss calculation})
\end{align}
\end{subequations}
In particular, stop gradient is applied to $\hat{\mathbf{x}}_0$ in the first step to prevent the gradient calculation across two steps, to reduce training time. Optionally, a model with exponential moving averaged weights can be used, but it requires even more memory. Compared to~\Cref{eq:ddpm-train-segmentation}, recycling modification only affects training and does not change network architecture. It is independent of the sampling strategy during inference. Therefore, the DDIM sampler can also be used for inference.

The recycling strategy we propose in this work differs from the one introduced in~\citet{fu2023importance} (denoted as ``Diff. rec. $\x_{t+1}$''), illustrated in~\Cref{fig:method_comparison} and the equations below,
\begin{subequations}\label{eq:ddpm-train-segmentation-recycling-next}
\begin{align}
    \x_{t+1} &\sim \mathcal{N}(\x_{t+1};\sqrt{\bar{\alpha}_{t+1}}\x_0, (1-\bar{\alpha}_t)\mathbf{I}),&(\text{rec.}~\x_{t+1},~\text{step 1, sampling}) \label{eq:rec-next-step1-sample}\\
    \hat{\mathbf{x}}_0 &= \text{StopGradient}(f_\theta(I,t+1,\mathbf{x}_{t+1})),&(\text{rec.}~\x_{t+1},~\text{step 1, prediction}) \label{eq:rec-next-step1-pred}\\
    \x_t &\sim \mathcal{N}(\x_t;\sqrt{\bar{\alpha}_t}\hat{\mathbf{x}}_0, (1-\bar{\alpha}_t)\mathbf{I}),&(\text{rec.}~\x_{t+1},~\text{step 2, sampling}) \label{eq:rec-next-step2-sample}\\
    \hat{\mathbf{x}}_0 &= f_\theta(I,t,\mathbf{x}_t),&(\text{rec.}~\x_{t+1},~\text{step 2, prediction}) \label{eq:rec-next-step2-pred}\\
    L_\text{denoising}(\theta)
    &=\E_{t,\x_0,\x_t}L(\x_{0},\hat{\mathbf{x}}_0),&(\text{loss calculation})
\end{align}
\end{subequations}
In the new approach (``Diff. rec. $\x_T$''), the first step is consistently executed at the time step $T$ instead of $t+1$ as shown in~\Cref{eq:ddpm-train-segmentation-recycling-max}. Compared to $\x_{t+1}$ in~\Cref{eq:ddpm-train-segmentation-recycling-next}, $x_T$ is fully noised and contains even less ground truth information during the initial step. Specifically, for a given time step $t$, $\x_t \sim \mathcal{N}(\x_t;\sqrt{\bar{\alpha}_t}\x_0, (1-\bar{\alpha}_t)\mathbf{I})$, which can be reparameterized as $\x_t=\sqrt{\bar{\alpha}_t}\x_0+\sqrt{1-\bar{\alpha}_t}\eps_t$ with $\eps_t\sim\mathcal{N}(\mathbf{0},\mathbf{I})$ and $\bar{\alpha}_t = \prod_{s=1}^t \alpha_s$. In this work, $\alpha_t$ is a monotonically decreasing noise schedule ranging from $0.999$ to $0.98$ for $t=1$ to $T$. Correspondingly, $\sqrt{\bar{\alpha}_t}$ monotonically decreases from $0.99995$ to $0.00632$. $\x_T=\sqrt{\bar{\alpha}_T}\x_0+\sqrt{1-\bar{\alpha}_T}\eps_T$ with $\sqrt{\bar{\alpha}_T}=0.00632$ can be considered to contain almost no ground truth information. The information can also be empirically measured by cross entropy and Dice score, and an example is presented in~\Cref{fig:x_t} in~\Cref{app:noise}. This seemingly minor modification removes the ground truth information from model inputs, essentially reducing the risk of data leakage and training overfitting. This adaptation guides the model to learn the denoising task based on its initial prediction, rather than ground truth. Consequently, the model can effectively denoise and refine the provided noisy mask, ultimately predicting the ground truth.

Recycling also differs from the self-conditioning methods proposed in~\citet{chen2022analog} (``Diff. sc. $\x_t$'') and~\citet{watson2023novo} (``Diff. sc. $\x_{t+1}$''). Although self-conditioning also requests two forward loops during training, it differs from recycling in multiple aspects. First, noisy samples in self-conditioning are always generated based on ground truth $\x_0$, while the second forward step of recycling does not rely on ground truth for noisy sample generation. Second, self-conditioning provides an additional input $\hat{\x}_0$, while recycling does not. Lastly, in self-conditioning, $\hat{\x}_0$ is replaced by zeros with $50\%$ probabilities, while recycling is applied constantly. The training strategy has been detailed in~\Cref{fig:method_comparison} and~\Cref{app:self-conditioning}. For further details, we refer the reader to the reference papers~\citep{chen2022analog,watson2023novo}.

\section{Experiments}
\subsection{Experiment Setting}
A range of experiments have been performed in four data sets (\Cref{subsec:data}) to evaluate the proposed method and the trained models from different aspects.

\subsubsection{Diffusion Training Strategy Comparison}
First, the proposed recycling training strategy (``Diff. rec. $\x_T$'') was compared with standard diffusion models (``Diff.'') and other diffusion training strategies that require two forward steps to evaluate the training efficiency with identical network architectures and compute budget. The compared diffusion training strategies include the previously proposed recycling method~\citet{fu2023importance} (``Diff. rec. $\x_{t+1}$'') and two self-conditioning techniques from~\citet{chen2022analog} (``Diff. sc. $\x_t$'') and~\citet{watson2023novo} (``Diff. sc. $\x_{t+1}$''). For each trained model using a different strategy, both DDPM and DDIM samplers were evaluated. Importantly, the predictions at each inference step were assessed to study the variation of performance along the inference process.

\subsubsection{Comparison to Non-diffusion Models}
The proposed methods were compared with non-diffusion-based models using identical architectures and the same compute budget. An ensemble model was also evaluated, where the predicted probabilities from the diffusion model and non-diffusion model were averaged. Models' segmentation accuracy was assessed with different granularities: per foreground class or averaged across foreground classes. Balnd-altmann plots were used to analyze the differences between models. 

\subsubsection{Ablation Studies for Recycling}
Ablation studies were performed, including assessing the performance with different lengths of inference and evaluating the stochasticity across different seeds during inference. Compared to the previous work~\citep{fu2023importance}, the effectiveness of the Transformer architecture and the change of training noise schedule was evaluated.

\subsubsection{Evaluation Metrics}
Different methods were evaluated using binary Dice score (DS) and $95\%$ Hausdorff distance (HD), averaging over foreground classes on the test sets. Dice score is reported in percentage, between $0\%$ and $100\%$. For Hausdorff distance, the values are in mm for 3D volumes and pixels for 2D images. Paired Student's t-tests with a significance level of $\alpha=0.05$ were performed on the Dice score to test statistical significance between model performance.

\subsection{Data}\label{subsec:data}
\subsubsection{Muscle Ultrasound}
The data set\footnote{\url{https://data.mendeley.com/data sets/3jykz7wz8d/1}}~\citep{marzola2021deep} provides
$3910$ labeled transverse musculoskeletal ultrasound images, which were split into $2531$, $666$, and $713$ images for
training, validation, and test sets, respectively. Images had the shape $480\times512$.
The predicted masks were post-processed, following~\citet{marzola2021deep}.
After filling the holes, multiple morphological operations were performed, including an erosion with a disk of radius 3 pixels, a dilation with a disk of radius 5 pixels, and an opening with a disk of radius 10 pixels. Afterward, only the largest connected component was preserved if the second largest structure was smaller than 75\% of the largest one; otherwise, 
the most superficial (i.e., towards the top of the image) one between the two largest components was preserved. Finally, holes were filled if there were any.

\subsubsection{Abdominal CT (AMOS)}
The data set\footnote{\url{https://zenodo.org/record/7155725\#.ZAkbe-zP2rO}}~\citep{ji2022amos} provides $200$ and $100$ CT image-mask pairs for $15$ abdominal organs in training and validation sets. The validation set was randomly split into non-overlapping validation and test sets, with $10$ and $90$ images, respectively. The images were first resampled with a voxel dimension
of $1.5\times1.5\times5.0$ (mm). HU values were clipped to $[-991, 362]$ and images were normalized so that the intensity had zero mean and unit variance. Lastly, images were center-cropped to shape $192\times128\times128$. During training, the patch size was $128\times128\times128$. During inference, the overlap between patches is $64\times0\times0$, and the predictions on the overlap were averaged.

\subsubsection{Prostate MR}
The data set\footnote{\url{https://zenodo.org/record/7013610\#.ZAkaXuzP2rM}}~\citep{li2022prototypical} contains $589$ T2-weighted image-mask pairs for $8$ anatomical structures from $7$ institutions. The images were randomly split into non-overlapping training, validation, and test sets, with $411$, $14$, and $164$ images in each split, respectively. The validation split has two images of each institution. The images were resampled with a voxel dimension
of $0.75\times0.75\times2.5$ (mm). Afterward, images were normalized so that the intensity had zero mean and unit variance. Lastly, the images were center-cropped
to shape $256\times256\times48$. During training, the patch size was $256\times256\times32$. During inference, the overlap between patches was $0\times0\times16$, and the predictions on the overlap were averaged.

\subsubsection{Brain MR (BraTS 2021)}
The data set\footnote{\url{https://www.kaggle.com/data sets/dschettler8845/brats-2021-task1}}~\citep{baid2021rsna} provides $1251$MR segmented mpMRI scans for brain tumour. The data set was randomly split into non-overlapping training, validation, and test sets, with $938$, $31$, and $282$ samples, respectively. The whole tumor mask was generated as foreground class, including GD-enhancing tumor, the peritumoral edematous/invaded tissue,
and the necrotic tumor core. Therefore, the task was a binary segmentation. Four modalities are available, including T1-weighted (T1), post-contrast T1-weighted (T1Gd), T2-weighted (T2), and T2 Fluid Attenuated Inversion Recovery (T2-FLAIR). The voxel dimension was $1.0\times1.0\times1.0$ (mm). Images were firstly normalized so that the intensity has zero mean and unit variance. Lastly, images were center-cropped to shape $179\times219\times155$ to remove the common background. During training, the patch size was $128\times128\times128$. During inference, the overlap between patches was $77\times37\times101$, and the predictions on the overlap were averaged.

\subsection{Implementation Details}
\label{subsec:implementation}
\begin{figure}[!ht]
\centering
\includegraphics[width=\linewidth]{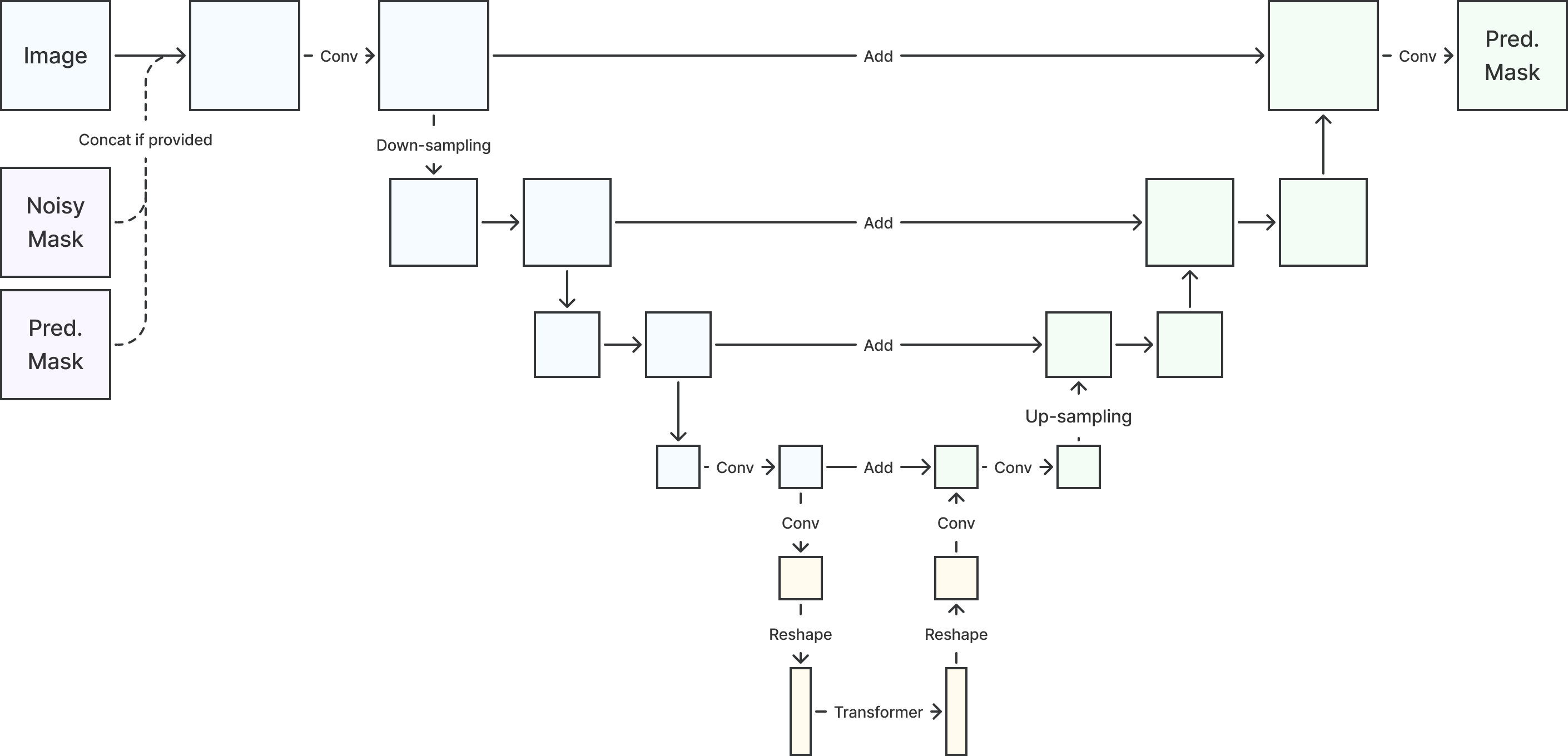}
\caption{\textbf{Unet architecture for diffusion and non-diffusion models.} The inputs are concatenated when a noisy mask (from diffusion models) or predicted mask (from self-conditioning) is provided. The tensor is enriched with convolution (time-conditioned for diffusion models) and down-sampling layers, then passed into a Transformer with positional encoding, the output is then enriched with convolution and up-sampling layers, and finally, prediction is performed with an additional $1\times1$ convolutional layer. ``Pred.'' stands for predicted.} \label{fig:unet}
\end{figure}
2D and 3D U-net variants with attention mechanisms were used for benchmarking the reference performance from cross-data-set non-diffusion models. The architecture is illustrated in~\Cref{fig:unet}. U-nets have four layers with $32$, $64$, $128$, and $256$ channels, respectively. The numbers of learnable parameters are summarized in~\Cref{tab:network-size} in~\Cref{app:implementation-details}. For diffusion-based models, the noise-corrupted masks were concatenated. Time was encoded using sinusoidal positional embedding~\citep{rombach2022high} and used in the convolution layers. 

For denoising training, a linear $\beta$ schedule between $0.0001$ and $0.02$ was used for $T=1001$ (illustrated in~\Cref{fig:x_t} in~\Cref{app:noise}). The segmentation-specific loss function is a weighted sum of cross-entropy and foreground-only Dice loss, with weight $20$ and $1$ respectively~\citep{kirillov2023segment}. Random rotation, translation, and scaling were adopted for data augmentation during training. Training hyper-parameters are listed in~\Cref{tab:training-hparams} in~\Cref{app:implementation-details}. Hyper-parameters were configured empirically without extensive tuning.

Models were trained once and checkpoints were saved every $500$ step. The checkpoint that had the best mean binary Dice score (without background class) in the validation set was used for the testing. For DDIM, the training was the same as DDPM while both validation and testing were performed using DDIM. The variance schedule was down-sampled to $5$ steps~\citep{nichol2021improved}. Experiments were carried out using bfloat16 mixed precision on TPU v3-8, which has $16\times8$ GB device memory. However, each device has only 16 GB memory, meaning that the model and data have to fit into $16$ GB memory. The JAX-based framework has been released on \url{https://github.com/mathpluscode/ImgX-DiffSeg}.

\section{Results and Discussion}

\subsection{Diffusion Training Strategy Comparison}
\begin{table}[!ht]
\centering
\caption{\textbf{Diffusion training strategies comparison.} ``Diff.'' represents standard diffusion. ``Diff. sc. $\x_{t}$'' and ``Diff. sc. $\x_{t+1}$'' represents self-conditioning from~\citet{chen2022analog} and~\citet{watson2023novo}, respectively. ``Diff. rec. $\x_{t+1}$'' and ``Diff. rec. $\x_T$'' represents recycling from~\citet{fu2023importance} and the proposed recycling in this work, respectively. The best results are in bold and underline indicates the difference to the second best is significant with p-value $<0.05$.}
\label{tab:diffusion_ablation}

\begin{subtable}[h]{\textwidth}
\centering
\begin{tabular}{l|cc|cc}
\toprule
\multirow{2}{*}{Method} & \multicolumn{2}{c|}{DDPM} & \multicolumn{2}{c}{DDIM} \\ \cline{2-5}
& DS $\uparrow$ & HD $\downarrow$ & DS $\uparrow$ & HD $\downarrow$ \\
\midrule
Diff. & 86.60 $\pm$ 12.38 & 41.11 $\pm$ 35.48 & 86.18 $\pm$ 12.41 & 42.31 $\pm$ 35.82 \\
Diff. sc. $\x_{t}$ & 86.35 $\pm$ 14.14 & 40.42 $\pm$ 37.53 & 85.96 $\pm$ 13.78 & 42.00 $\pm$ 36.76 \\
Diff. sc. $\x_{t+1}$ & 87.14 $\pm$ 11.48 & 39.24 $\pm$ 32.83 & 86.30 $\pm$ 11.49 & 41.89 $\pm$ 32.72 \\
Diff. rec. $\x_{t+1}$ & 87.44 $\pm$ 12.39 & 39.68 $\pm$ 36.21 & 87.43 $\pm$ 12.25 & 39.82 $\pm$ 35.39 \\
Diff. rec. $\x_{T}$ & \textbf{\underline{88.23 $\pm$ 11.69}} & \textbf{\underline{35.37 $\pm$ 31.79}} & \textbf{\underline{88.21 $\pm$ 11.70}} & \textbf{\underline{35.52 $\pm$ 31.91}} \\
\bottomrule
\end{tabular}
\caption{Muscle Ultrasound}
\label{tab:diffusion_ablation_muscle}
\end{subtable}

\begin{subtable}[h]{\textwidth}
\centering
\begin{tabular}{l|cc|cc}
\toprule
\multirow{2}{*}{Method} & \multicolumn{2}{c|}{DDPM} & \multicolumn{2}{c}{DDIM} \\ \cline{2-5}
& DS $\uparrow$ & HD $\downarrow$ & DS $\uparrow$ & HD $\downarrow$ \\
\midrule
Diff. & 85.25 $\pm$ 5.36 & 7.12 $\pm$ 3.83 & 85.59 $\pm$ 5.24 & 7.13 $\pm$ 3.98 \\
Diff. sc. $\x_{t}$ & 86.04 $\pm$ 5.12 & 7.06 $\pm$ 4.20 & 85.50 $\pm$ 5.14 & 7.21 $\pm$ 4.16 \\
Diff. sc. $\x_{t+1}$ & 85.86 $\pm$ 5.27 & 6.98 $\pm$ 3.54 & 85.25 $\pm$ 5.42 & 7.28 $\pm$ 3.72 \\
Diff. rec. $\x_{t+1}$ & 86.48 $\pm$ 5.24 & 6.69 $\pm$ 4.59 & 86.35 $\pm$ 5.31 & 6.75 $\pm$ 4.55 \\
Diff. rec. $\x_{T}$ & \textbf{\underline{87.45 $\pm$ 5.43}} & \textbf{6.56 $\pm$ 5.44} & \textbf{\underline{87.45 $\pm$ 5.43}} & \textbf{6.55 $\pm$ 5.43} \\
\bottomrule
\end{tabular}
\caption{Abdominal CT}
\label{tab:diffusion_ablation_abdominal}
\end{subtable}

\begin{subtable}[h]{\textwidth}
\centering
\begin{tabular}{l|cc|cc}
\toprule
\multirow{2}{*}{Method} & \multicolumn{2}{c|}{DDPM} & \multicolumn{2}{c}{DDIM} \\ \cline{2-5}
& DS $\uparrow$ & HD $\downarrow$ & DS $\uparrow$ & HD $\downarrow$ \\
\midrule
Diff. & 83.61 $\pm$ 4.87 & 5.10 $\pm$ 2.40 & 83.11 $\pm$ 4.81 & 5.00 $\pm$ 2.35 \\
Diff. sc. $\x_{t}$ & 83.47 $\pm$ 4.85 & 5.17 $\pm$ 2.65 & 82.49 $\pm$ 4.88 & 5.42 $\pm$ 2.70 \\
Diff. sc. $\x_{t+1}$ & 83.97 $\pm$ 4.85 & 4.93 $\pm$ 2.66 & 83.00 $\pm$ 4.89 & 5.10 $\pm$ 2.64 \\
Diff. rec. $\x_{t+1}$ & 84.29 $\pm$ 5.12 & 4.59 $\pm$ 2.21 & 84.21 $\pm$ 4.89 & 4.96 $\pm$ 2.92 \\
Diff. rec. $\x_{T}$ & \textbf{\underline{85.54 $\pm$ 5.20}} & \textbf{4.40 $\pm$ 1.96} & \textbf{\underline{85.54 $\pm$ 5.20}} & \textbf{\underline{4.41 $\pm$ 1.96}} \\
\bottomrule
\end{tabular}
\caption{Prostate MR}
\label{tab:diffusion_ablation_prostate}
\end{subtable}

\begin{subtable}[h]{\textwidth}
\centering
\begin{tabular}{l|cc|cc}
\toprule
\multirow{2}{*}{Method} & \multicolumn{2}{c|}{DDPM} & \multicolumn{2}{c}{DDIM} \\ \cline{2-5}
& DS $\uparrow$ & HD $\downarrow$ & DS $\uparrow$ & HD $\downarrow$ \\
\midrule
Diff. & 90.29 $\pm$ 12.98 & 8.46 $\pm$ 15.55 & 89.94 $\pm$ 13.00 & 8.55 $\pm$ 15.50 \\
Diff. sc. $\x_{t}$ & 90.12 $\pm$ 12.39 & 9.55 $\pm$ 17.18 & 89.73 $\pm$ 12.61 & 9.67 $\pm$ 16.86 \\
Diff. sc. $\x_{t+1}$ & 89.11 $\pm$ 14.70 & 9.63 $\pm$ 17.47 & 88.75 $\pm$ 14.77 & 9.62 $\pm$ 16.97 \\
Diff. rec. $\x_{t+1}$ & 86.97 $\pm$ 10.94 & 9.83 $\pm$ 12.62 & 84.76 $\pm$ 13.42 & 12.52 $\pm$ 15.55 \\
Diff. rec. $\x_{T}$ & \textbf{\underline{92.29 $\pm$ 8.55}} & \textbf{7.03 $\pm$ 13.48} & \textbf{\underline{92.29 $\pm$ 8.55}} & \textbf{\underline{7.03 $\pm$ 13.48}} \\
\bottomrule
\end{tabular}
\caption{Brain MR}
\label{tab:diffusion_ablation_brain}
\end{subtable}

\end{table}

Our proposed recycling method (Diff. rec. $\x_T$) achieved mean Dice scores of $88.23\%$, $87.45\%$, $85.54\%$, and $92.29\%$ on muscle ultrasound, abdominal CT, prostate MR, and brain MR data sets, respectively. These scores marked absolute improvements of $1.63\%$, $2.20\%$, $1.93\%$, and $2.00\%$ over standard diffusion models, respectively. The relative improvements are $1.88\%$, $2.58\%$, $2.31\%$, and $2.22\%$ respectively. Impressively, this novel strategy consistently outperformed the other three training approaches in terms of both Dice score and Hausdorff distance. The observed differences were significant for all data sets in terms of Dice score ($p=0.003$ for muscle ultrasound and $p<0.001$ for other data sets). These findings held for both the DDPM and the DDIM samplers, underscoring the wide applicability of the proposed training strategy.
\afterpage{
\begin{figure}[!ht]
     \centering
     \begin{subfigure}[b]{0.45\textwidth}
         \centering
         \includegraphics[width=\textwidth]{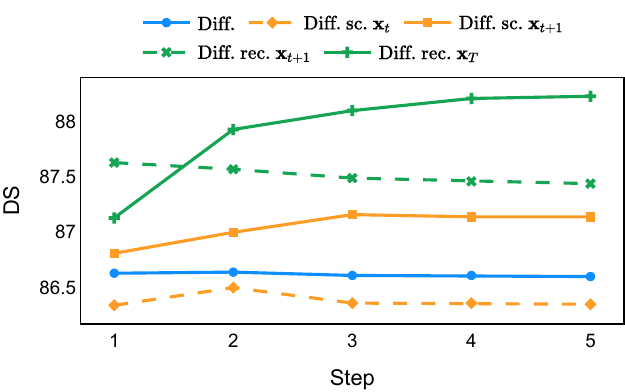}
         \caption{Dice Score for muscle ultrasound}
         \label{fig:per_step_muscle_us_DDPM_DS}
     \end{subfigure}
     \begin{subfigure}[b]{0.45\textwidth}
         \centering
         \includegraphics[width=\textwidth]{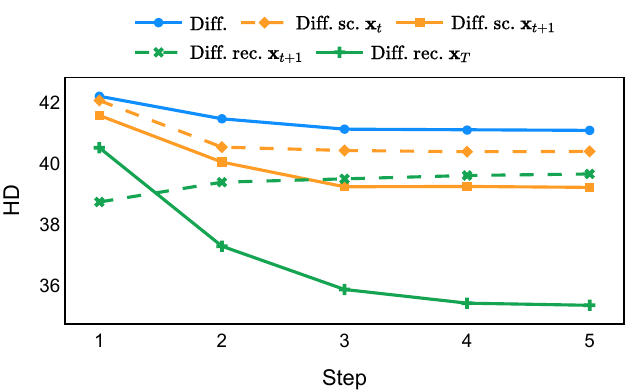}
         \caption{Hausdorff distance for muscle ultrasound}
         \label{fig:per_step_muscle_us_DDPM_HD}
     \end{subfigure}
     \begin{subfigure}[b]{0.45\textwidth}
         \centering
         \includegraphics[width=\textwidth]{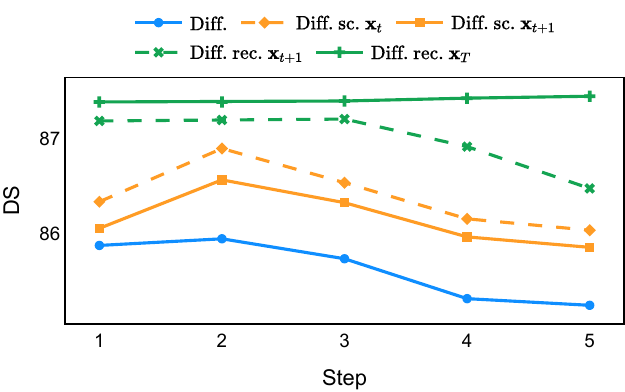}
         \caption{Dice score for abdominal CT}
         \label{fig:per_step_amos_ct_DDPM_DS}
     \end{subfigure}
     \begin{subfigure}[b]{0.45\textwidth}
         \centering
         \includegraphics[width=\textwidth]{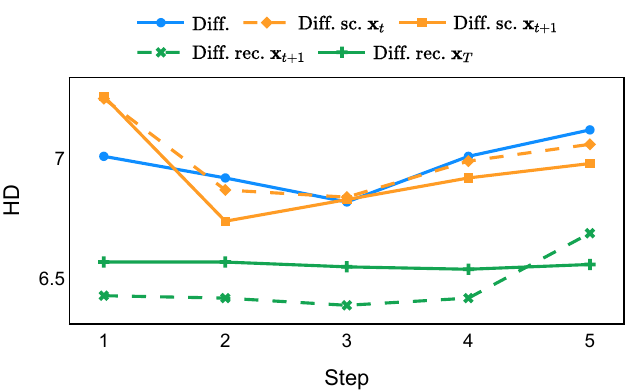}
         \caption{Hausdorff distance for abdominal CT}
         \label{fig:per_step_amos_ct_DDPM_HD}
     \end{subfigure}
     \begin{subfigure}[b]{0.45\textwidth}
         \centering
         \includegraphics[width=\textwidth]{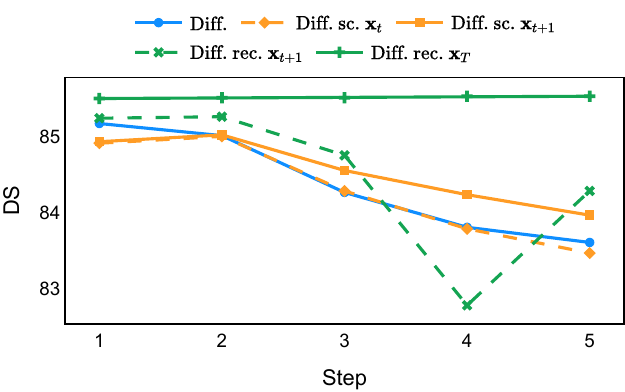}
         \caption{Dice score for prostate MR}
         \label{fig:per_step_male_pelvic_mr_DDPM_DS}
     \end{subfigure}
     \begin{subfigure}[b]{0.45\textwidth}
         \centering
         \includegraphics[width=\textwidth]{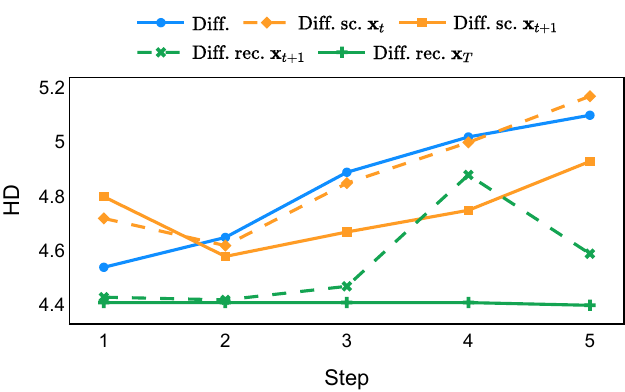}
         \caption{Hausdorff distance for prostate MR}
         \label{fig:per_step_male_pelvic_mr_DDPM_HD}
     \end{subfigure}
     \begin{subfigure}[b]{0.45\textwidth}
         \centering
         \includegraphics[width=\textwidth]{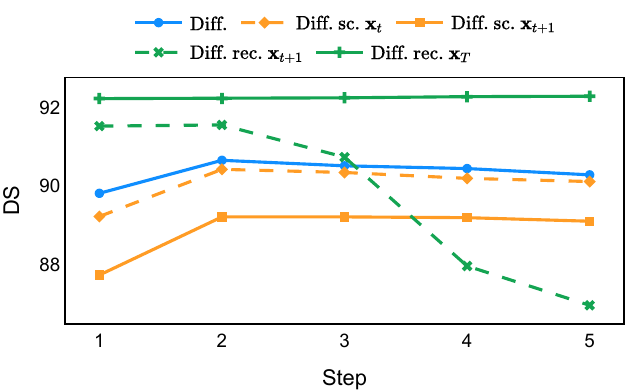}
         \caption{Dice score for brain MR}
         \label{fig:per_step_brats2021_mr_DDPM_DS}
     \end{subfigure}
     \begin{subfigure}[b]{0.45\textwidth}
         \centering
         \includegraphics[width=\textwidth]{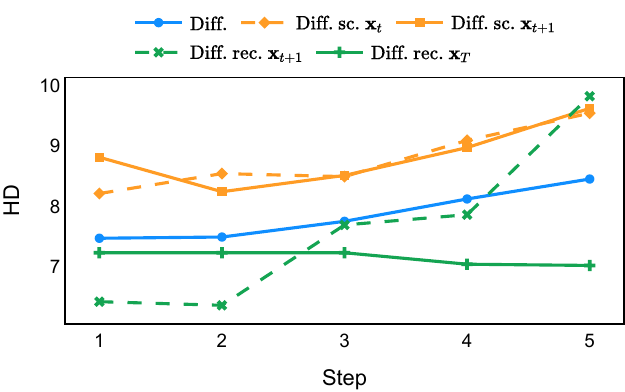}
         \caption{Hausdorff distance for brain MR}
         \label{fig:per_step_brats2021_mr_DDPM_HD}
     \end{subfigure}
    \caption{\textbf{Segmentation performance per step.} ``Diff.'' represents standard diffusion. ``Diff. sc. $\x_{t}$'' and ``Diff. sc. $\x_{t+1}$'' represents self-conditioning from~\citet{chen2022analog} and~\citet{watson2023novo}, respectively. ``Diff. rec. $\x_{t+1}$'' and ``Diff. rec. $\x_T$'' represents recycling from~\citet{fu2023importance} and the proposed recycling in this work, respectively. The sampler is DDPM.}
    \label{fig:per_step_ddpm}
\end{figure}
\clearpage
}
\afterpage{
\begin{figure}[!ht]
\centering
\includegraphics[width=\textwidth]{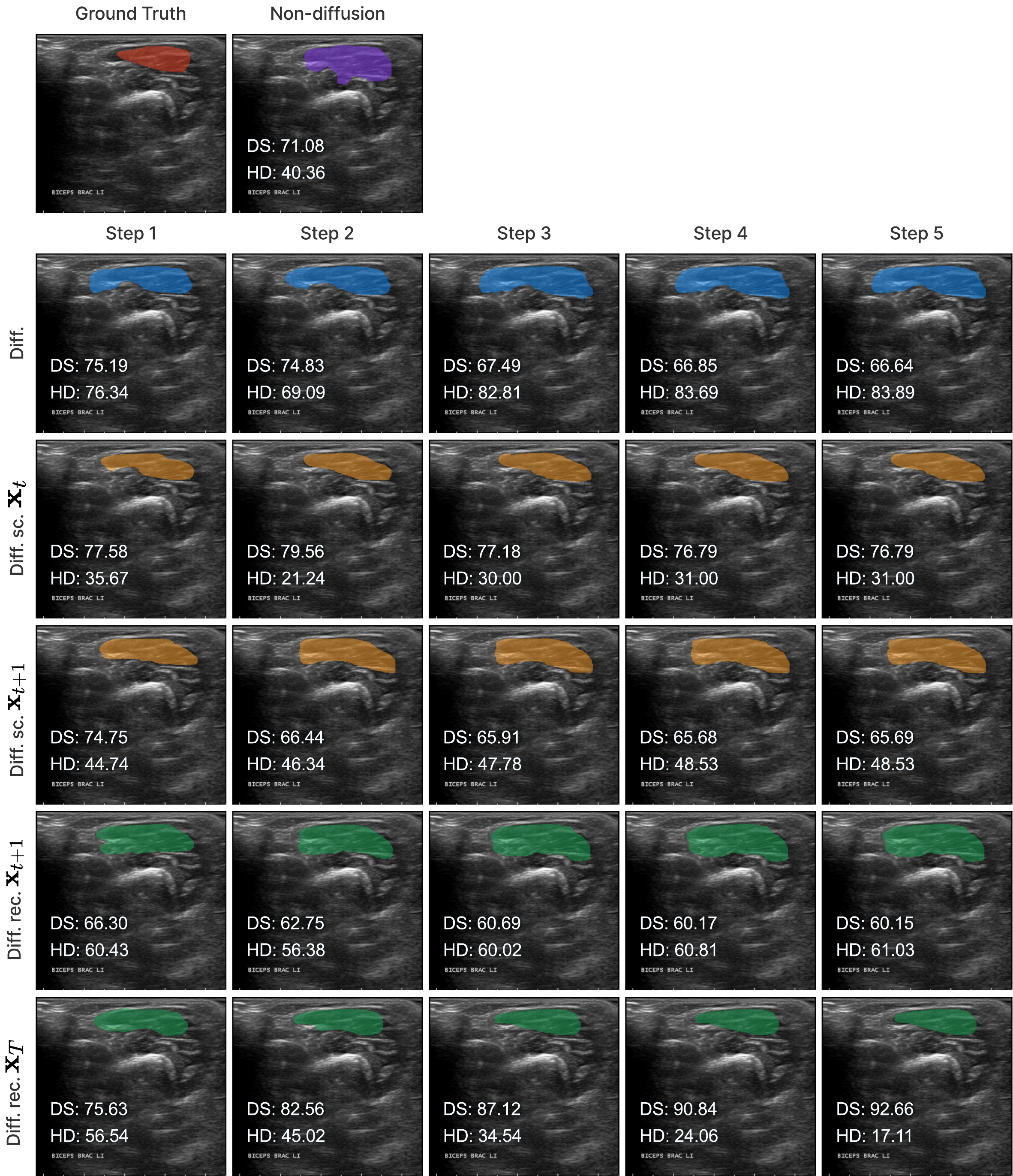}
\caption{\textbf{Diffusion training strategies comparison on muscle ultrasound example.} ``Diff.'' represents standard diffusion. ``Diff. sc. $\x_{t}$'' and ``Diff. sc. $\x_{t+1}$'' represents self-conditioning from~\citet{chen2022analog} and~\citet{watson2023novo}, respectively. ``Diff. rec. $\x_{t+1}$'' and ``Diff. rec. $\x_T$'' represents recycling from~\citet{fu2023importance} and the proposed recycling in this work, respectively. The Dice score (DS) and Hausdorff distance (HD) for each sample are labeled at the bottom. While different diffusion models have similar performance on the first step, the proposed method (last row) can refine the segmentation mask.}
\label{fig:comparison_muscle_ultrasound}
\end{figure}
\clearpage
}

As depicted in~\Cref{fig:first_last_step_diff_ddpm} in~\Cref{app:results-diff}, standard diffusion models often produce segmentation masks in the last step that are less accurate than the initial prediction. Similar challenges were observed with self-conditioning strategies and previously proposed recycling methods. The newly introduced recycling method was the only approach that improved initial segmentation predictions for more than half of the test images. Moreover, the average performance per step has been visualized in~\Cref{fig:per_step_ddpm}, where diffusion models frequently exhibit gradually declining or unstable performance during inference, in terms of both Dice score and Hausdorff distance. It is interesting to observe that often the optimal prediction emerges not at the final step but rather at an intermediate stage. This has been observed in all diffusion models except the newly proposed diffusion model with the innovative recycling method. In the latter case, the quality of segmentation consistently improved or remained stable throughout the inference process, distinguishing it from the observed trend. A qualitative comparison on an example muscle ultrasound image has been illustrated in~\Cref{fig:comparison_muscle_ultrasound}, where the proposed diffusion model was able to refine the segmentation mask progressively. Similar observations have been noted with the DDIM sampler as well, as shown in~\Cref{fig:first_last_step_diff_ddim} and~\Cref{fig:per_step_ddim}. This finding aligns with the discussions from~\citet{kolbeinsson2022multi,lai2023denoising} that the diffusion-based segmentation model performance is strongly influenced by the prediction of the initial step. For self-conditioning or the previously proposed recycling, the denoising training relies on the ground truth to varying degrees therefore the diffusion models are trained with ground truth-like initial predictions. However, no ground truth is available during inference, and the distributions of initial predictions from the trained models are dissimilar from ground truths. This results in an out-of-sample inference and therefore a declining performance. In contrast, the proposed method ingests model predictions for both the training and inference phases without the bias toward ground truth. These observations reaffirm the importance and benefits of harmonizing the training and inference processes. This alignment is crucial to mitigate data leakage, prevent overfitting, and help generalization.

\subsection{Comparison to Non-diffusion Models}
\begin{table}[!ht] 
\centering
\caption{\textbf{Segmentation performance comparison to non-diffusion models.} ``No diff.'' represents non-diffusion model. ``Diff. rec. $\x_T$'' represents the diffusion model with proposed recycling. ``Ensemble'' represents the model averaging the probabilities from ``No diff.'' and ``Diff. rec. $\x_T$''. The inference sampler is DDPM. The best results are in bold and underline indicates the difference to non-diffusion model is significant with p-value < $0.05$.}
\label{tab:non_diffusion_ablation}

\begin{tabular}{l|l|c|c}
\toprule
Data Set & Method & DS $\uparrow$ & HD $\downarrow$ \\
\midrule
\multirow{3}{*}{Muscle Ultrasound} 
& No diff. & 88.15 $\pm$ 10.77 & 36.86 $\pm$ 30.04 \\
& Diff. rec. $\x_{T}$ & 88.23 $\pm$ 11.69 & 35.37 $\pm$ 31.79 \\
& Ensemble & \textbf{\underline{88.88 $\pm$ 10.59}} & \textbf{\underline{34.01 $\pm$ 28.75}} \\
\hline
\multirow{3}{*}{Abdominal CT} 
& No diff. & 87.59 $\pm$ 5.10 & 6.36 $\pm$ 3.86 \\
& Diff. rec. $\x_{T}$ & 87.45 $\pm$ 5.43 & 6.56 $\pm$ 5.44 \\
& Ensemble & \textbf{\underline{88.29 $\pm$ 5.21}} & \textbf{\underline{5.60 $\pm$ 3.13}} \\
\hline
\multirow{3}{*}{Prostate MR} 
& No diff. & 85.22 $\pm$ 5.18 & 4.62 $\pm$ 2.37 \\
& Diff. rec. $\x_{T}$ & \underline{85.54 $\pm$ 5.20} & 4.40 $\pm$ 1.96 \\
& Ensemble & \textbf{\underline{85.95 $\pm$ 5.12}} & \textbf{\underline{4.32 $\pm$ 2.01}} \\
\hline
\multirow{3}{*}{Brain MR} 
& No diff. & 92.43 $\pm$ 9.10 & 5.20 $\pm$ 9.56 \\
& Diff. rec. $\x_{T}$ & 92.29 $\pm$ 8.55 & \underline{7.03 $\pm$ 13.48} \\
& Ensemble & \textbf{\underline{92.67 $\pm$ 8.60}} & \textbf{5.03 $\pm$ 8.41} \\
\bottomrule
\end{tabular}
\end{table}
\afterpage{
\begin{figure}[!ht]
     \centering
     \includegraphics[width=\textwidth]{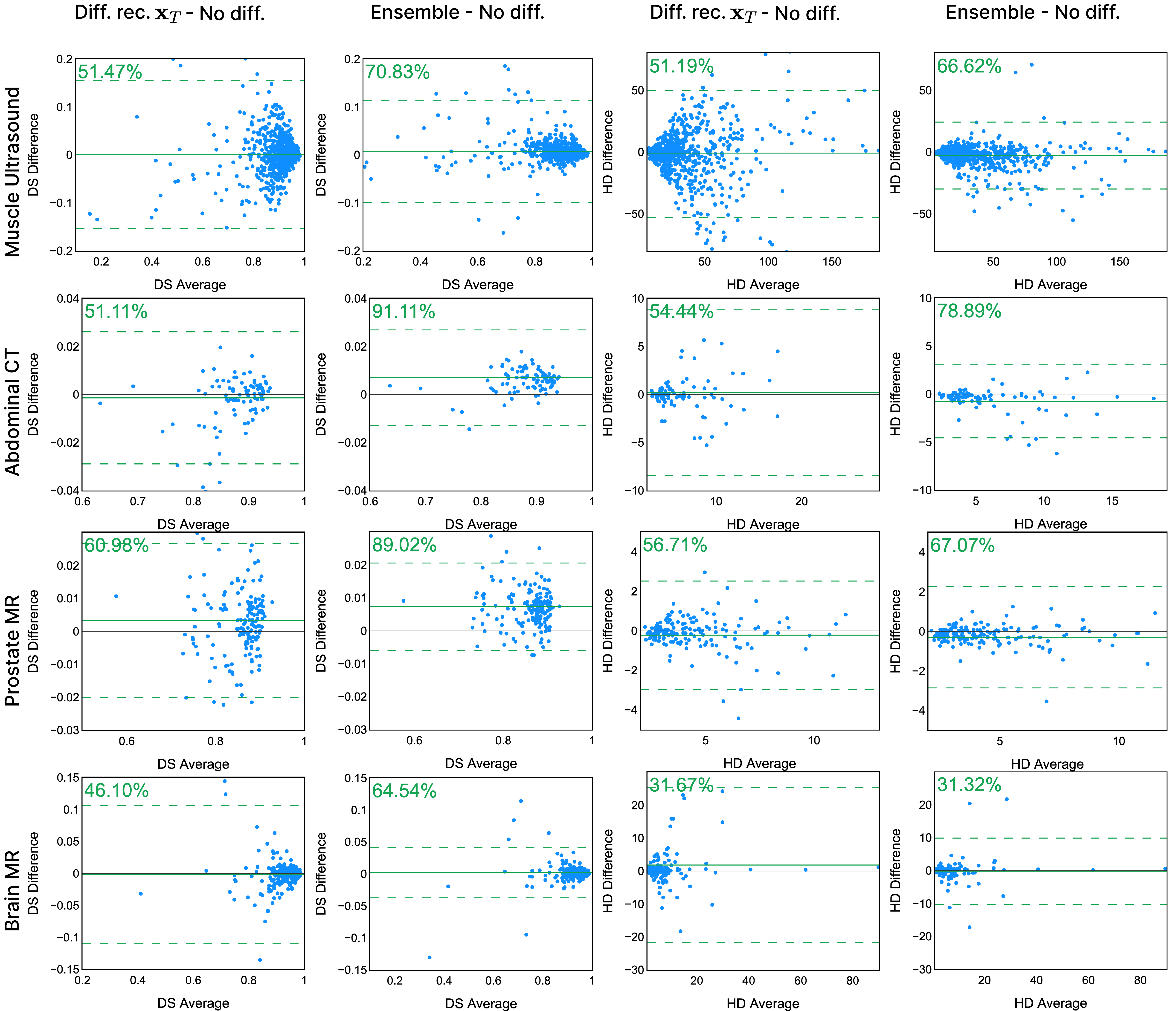}
    \caption{
    \textbf{Balnd-altmann plot for comparison of diffusion and ensemble models against non-diffusion models.} ``No diff.'' represents non-diffusion model. ``Diff. rec. $\x_T$'' represents the diffusion model with proposed recycling. ``Ensemble'' represents ensembled model by averaging predicted probabilities. The inference sampler is DDPM. DS and HD represents Dice score and Hausdorff distance, respectively. The differences are calculated against non-diffusion models. Positive dice score difference and negative Hausdorff distances indicate improvements. The green solid lines indicates the average difference and the dash lines are mean $\pm 1.96$ standard deviation of the difference. The percentage indicates the number of samples having better performance against non-diffusion baseline. Ensemble models brings an improvement of Dice score for $18.44\% - 40.00\%$ samples across applications.
    }
    \label{fig:diff_vs_nodiff}
\end{figure}
\begin{figure}[!ht]
     \centering
     \begin{subfigure}[b]{\textwidth}
         \centering
         \includegraphics[width=0.7\textwidth]{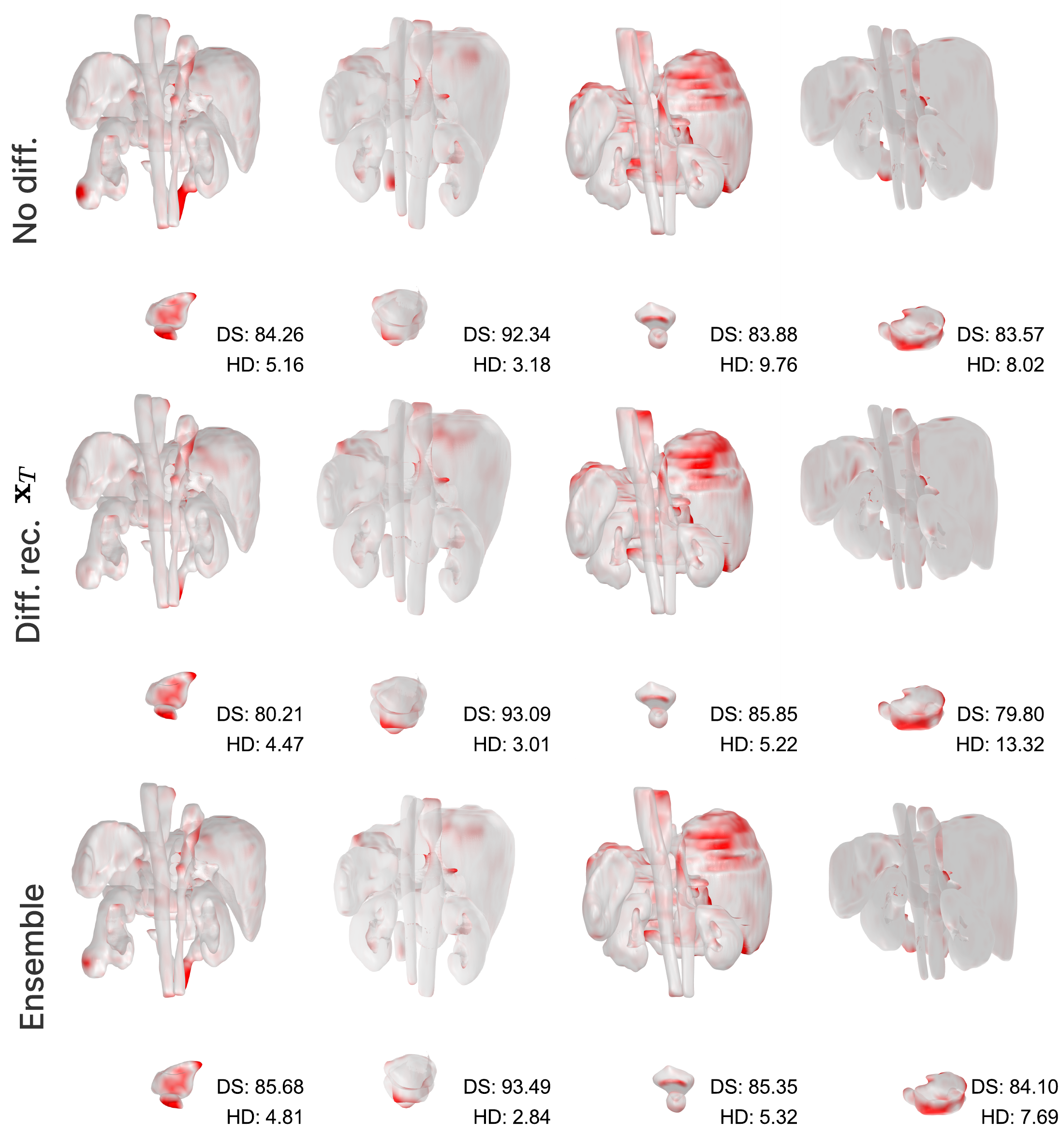}
         \caption{Structures in abdominal CT.}
         \label{fig:amos_ct_3d}
     \end{subfigure}
     \begin{subfigure}[b]{\textwidth}
         \centering
         \includegraphics[width=0.7\textwidth]{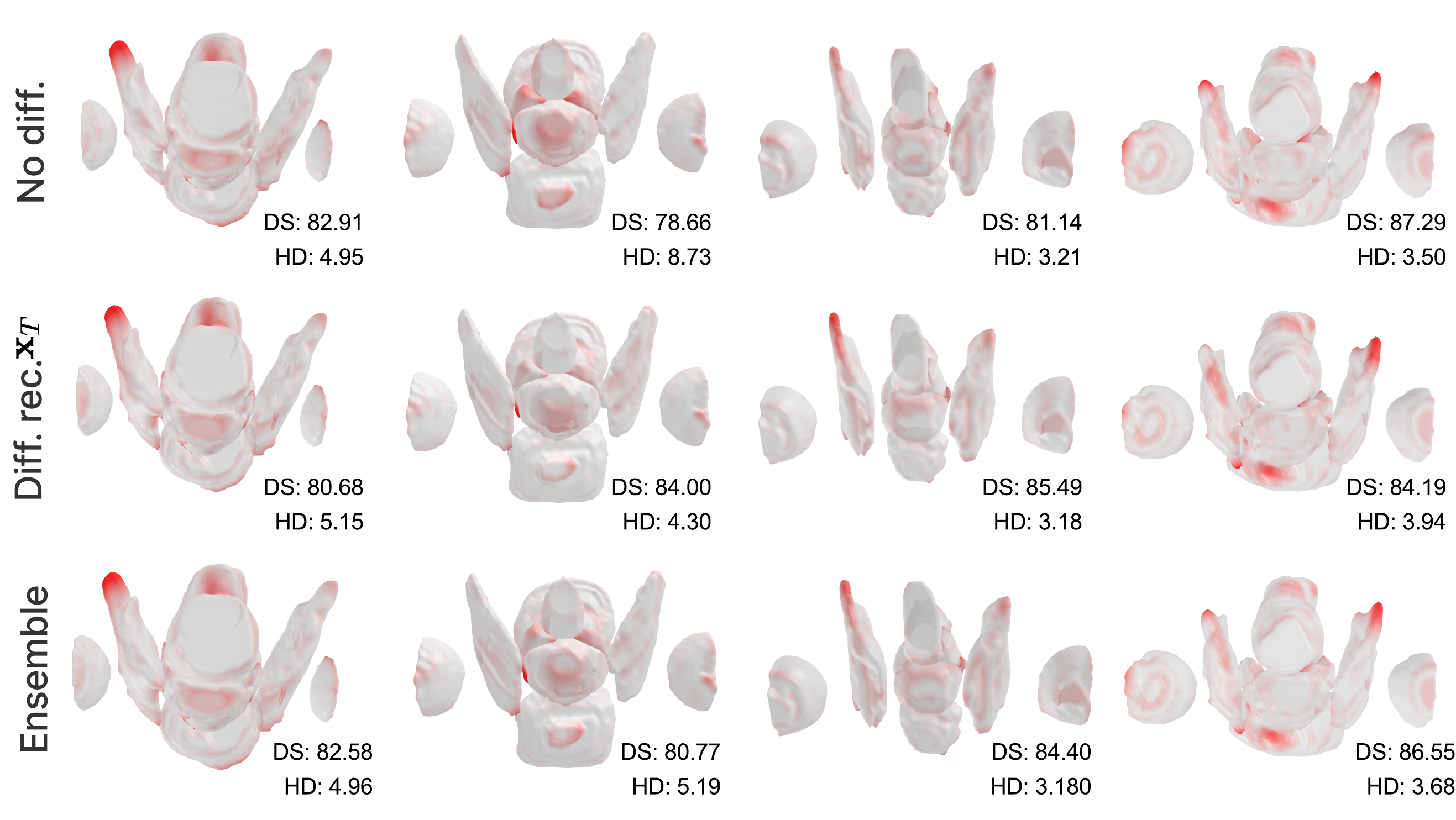}
         \caption{Structures in prostate MR.}
         \label{fig:prostate_mr_3d}
     \end{subfigure}
    \caption{\textbf{Segmentation error of non-diffusion-based and diffusion-based models.} ``No diff.'' represents non-diffusion model. ``Diff. rec. $\x_T$'' represents the diffusion model with proposed recycling. ``Ensemble'' represents ensembled model by averaging predicted probabilities. The ground truth segmentation is visualised. For each point on the surface, the distance to the surface of predicted segmentation is calculated and displayed with red color. The Dice score (DS) and Hausdorff distance (HD) for each sample are labeled at bottom.}
    \label{fig:3d}
\end{figure}
\clearpage
}
The proposed diffusion models (``Diff. rec. $\x_T$'') were compared with their non-diffusion counterparts (``No diff.''), where models with identical architectures were trained under the same scheme with the same compute budget. This provides a fair comparison without application-specific adjustments. For diffusion models, the performance with DDPM was selected. As shown in~\Cref{tab:non_diffusion_ablation}, The diffusion models yielded similar performance across all data sets. The difference in Dice score is not significant for muscle ultrasound, abdominal CT, and brain MR, but the diffusion model had a higher Dice score for prostate MR ($p = 0.001$). Furthermore, \Cref{fig:diff_vs_nodiff} shows that the proposed diffusion model achieved a higher Dice score on more than $50\%$ samples for muscle ultrasound, abdominal CT, and prostate MR data sets. To the best of our knowledge, this is the first time that diffusion models achieved comparable performance against standard non-diffusion-based models with the same architecture and compute budget.

By ensembling these two models via averaging the probabilities, we achieved mean Dice scores of $88.88\%$, $88.29\%$, $85.95\%$, and $92.67\%$ on muscle ultrasound, abdominal CT, prostate MR, and brain MR data sets, respectively. The improvements in Dice score were significant across all four data sets ($p = 0.037$ for brain MR and $p<0.001$ for other data sets). Especially, \Cref{fig:diff_vs_nodiff} shows that the ensemble model reached a higher Dice score compared to non-diffusion models on $70.83\%$, $91.11\%$, $89.02\%$, and $64.54\%$ samples in the test set for muscle ultrasound, abdominal CT, prostate MR data and brain MR, respectively. These scores marked an absolute increase of $19.36\%$, $40.00\%$, $28.04\%$, and $19.44\%$ compared to the diffusion model alone. Moreover, Abdominal CT and prostate MR are two data sets with multiple classes and their per-class segmentation performances are summarised in~\Cref{tab:per_class_dice} and~\Cref{tab:per_class_hausdorff} in~\Cref{app:results-diff}, respectively. Upon comparing diffusion models and non-diffusion models, neither consistently outperformed the other across all classes. However, the ensemble model reached the best performance across all classes and the improvement of Dice score is significant for 13 out of 15 classes in Abdominal CT data and all classes in prostate MR data (all p-values $<= 0.01$, excluding Spleen $p=0.06$ and Gall bladder $p=0.876$). Multiple examples have also been visualized in~\Cref{fig:3d} and~\Cref{fig:brain_mr_3d} for the segmentation error.

We highlight that the value of the competitive performance from alternative methods, in particular a different class of generative model-based approaches, is beyond the replacement of current segmentation algorithms for specific potential applications. Our results demonstrate a consistent improvement by combining diffusion and non-diffusion models across applications, even when they yielded a similar performance individually. This is one of the possible potential uses of the proposed improved diffusion models in addition to the well-established non-diffusion baseline. Future research could explore application-specific tuning for further performance improvements.

\subsection{Ablation Studies}
\subsubsection{Number of sampling steps}
\begin{table}[!ht]
\centering
\caption{\textbf{Diffusion with different number of sampling steps.} Sampler is DDPM. Diffusion models were trained using the proposed recycling method (Diff. rec. $\x_T$). OOM indicates that out of memory errors were encountered. Best results are in bold.}
\label{tab:test_timesteps_ablation}

\centering
\begin{tabular}{l|c|cc}
\toprule
Data Set & \# Sampling Steps & Dice Score & Hausdorff Distance \\
\midrule
\multirow{3}{*}{Muscle Ultrasound} 
& 2 & 88.01 $\pm$ 12.07 & 36.55 $\pm$ 32.66 \\
& 5 & 88.23 $\pm$ 11.69 & 35.37 $\pm$ 31.79 \\
& 11 & \textbf{88.30 $\pm$ 11.29} & \textbf{35.25 $\pm$ 30.64} \\
\midrule
\multirow{3}{*}{Abdominal CT} 
& 2 & 87.44 $\pm$ 5.43 & \textbf{6.56 $\pm$ 5.42} \\
& 5 & \textbf{87.45 $\pm$ 5.43} & 6.56 $\pm$ 5.44 \\
& 11 & OOM & OOM \\
\midrule
\multirow{3}{*}{Prostate MR} 
& 2 & 85.54 $\pm$ 5.19 & 4.40 $\pm$ 1.96 \\
& 5 & \textbf{85.54 $\pm$ 5.20} & \textbf{4.40 $\pm$ 1.96} \\
& 11 & \textbf{85.54 $\pm$ 5.20} & \textbf{4.40 $\pm$ 1.96} \\
\midrule
\multirow{3}{*}{Brain MR} 
& 2 & 92.29 $\pm$ 8.54 & 7.03 $\pm$ 13.47 \\
& 5 & \textbf{92.29 $\pm$ 8.55} & 7.03 $\pm$ 13.48 \\
& 11 & 92.29 $\pm$ 8.57 & \textbf{7.02 $\pm$ 13.48} \\
\bottomrule
\end{tabular}

\end{table}

Diffusion models were trained using a thousand steps, yet employing the same number of steps for inference can be cost-prohibitive, particularly for processing 3D image volumes. As a result, practical inference commonly utilizes a condensed schedule with a limited number of steps. While this approach reduces computational expenses, the resulting sample quality might be compromised. An ablation study of the numbers of timesteps during inference has therefore been performed across data sets with the proposed recycling-based diffusion model. DDPM sampler was used. The results have been summarised in~\Cref{tab:test_timesteps_ablation}. Notably, increasing the number of steps yielded a higher Dice score for the muscle ultrasound dataset but the difference is not significant ($p>=0.05$). For prostate MR and brain MR data sets, the models maintained almost the same performance regardless of the inference length ($p>=0.05$). Given that longer inference times and increased device memory usage are associated with more timesteps (e.g. out-of-memory errors were encountered with Abdominal CT at 11 steps), the trade-off between computational resources and performance suggests that a five-step sampling schedule provides the optimal balance.

\subsubsection{Inference Variance}
\begin{table}[!ht]
\centering
\caption{\textbf{Diffusion model performance across different inference seeds.} For each sample, the maximum difference ($\Delta$) across five random seeds is calculated. The average across all samples is reported.}
\label{tab:seed_ablation}

\begin{tabular}{c|ccccc}
\toprule
\multirow{2}{*}{Data Set} & \multicolumn{5}{c}{Mean $\Delta$ Dice Score} \\ \cline{2-6}
 & Step 1 & Step 2 & Step 3 & Step 4 & Step 5 \\
\midrule
Muscle Ultrasound & 0.0212 & 0.0165 & 0.0122 & 0.0081 & 0.0051 \\
Abdominal CT & 0.0009 & 0.0010 & 0.0009 & 0.0008 & 0.0004 \\
Prostate MR & 0.0004 & 0.0004 & 0.0004 & 0.0004 & 0.0002 \\
Brain MR & 0.0005 & 0.0005 & 0.0005 & 0.0003 & 0.0001 \\
\midrule
\multirow{2}{*}{Data Set} & \multicolumn{5}{c}{Mean $\Delta$ Hausdorff Distance} \\ \cline{2-6}
 & Step 1 & Step 2 & Step 3 & Step 4 & Step 5 \\
\midrule
Muscle Ultrasound & 10.0582 & 7.0020 & 4.7440 & 3.1758 & 1.8164 \\
Abdominal CT & 0.1481 & 0.1339 & 0.1221 & 0.0751 & 0.0673 \\
Prostate MR & 0.0447 & 0.0426 & 0.0499 & 0.0431 & 0.0209 \\
Brain MR & 0.0758 & 0.0779 & 0.0678 & 0.0616 & 0.0197 \\
\bottomrule
\end{tabular}

\end{table}

Different from deterministic models, the inference process of the diffusion model inherently incorporates stochasticity and models a distribution of the segmentation masks. Using the DDPM sampler with the proposed recycling-based diffusion model, the inference on each data set has been repeated with five different random seeds. Consequently, each sample has five distinct predicted masks. The maximum differences across five predictions were computed for the Dice score and Hausdorff distance, denoted by $\Delta$ Dice score and $\Delta$ Hausdorff distance, respectively. The average of this performance difference across all samples in the test set has been reported in~\Cref{tab:seed_ablation} for all data sets. While the magnitude of the average difference (mean $\Delta$) varies across data sets, a common trend was observed where mean $\Delta$ diminished during the sampling process for both metrics. In other words, despite different initial predictions, the model's predictions gradually converge as the difference across seeds decreases. Moreover, the relative magnitude of the mean $\Delta$ Hausdorff distance (e.g. $1.82$ at the last step for muscle ultrasound represents around $5\%$ fluctuation compared to $35.37$, the mean Hausdorff distance to ground truth) was larger than the relative magnitude for Dice score (e.g. $0.0051$ at the last step for muscle ultrasound was around $0.006\%$ fluctuation compared to $88.23$ the mean Hausdorff distance to ground truth). We hypothesize that the variation among predictions may predominantly revolve around local refinements in mask boundaries, as opposed to significant alterations like expansion or contraction of foreground areas. This may open a direction for further improving diffusion training: instead of performing independent noising per pixel/voxel results in fragmented and disjointed masks, the noising can be morphology-informed such that the noise-corrupted masks expand or contract the foreground with continuous boundaries.

\subsubsection{Transformer}
\begin{table}[!ht] 
\centering
\caption{\textbf{Segmentation performance without Transformer.} ``No diff.'' represents non-diffusion model. ``Diff. rec. $\x_T$'' represents the diffusion model with proposed recycling. The inference sampler is DDPM. 
The best results are in bold and underline indicates the difference to non-diffusion model is significant with p-value < $0.05$.
}
\label{tab:transformer_ablation}

\begin{tabular}{l|l|c|cc}
\toprule
Data Set & Method & Transformer & DS $\uparrow$ & HD $\downarrow$ \\
\midrule
\multirow{4}{*}{Muscle US} 
& \multirow{2}{*}{No diff.} & & 86.66 $\pm$ 13.16 & 45.01 $\pm$ 38.86 \\
& & \checkmark & \textbf{\underline{88.15 $\pm$ 10.77}} & \textbf{\underline{36.86 $\pm$ 30.04}} \\ \cline{2-5}
 & \multirow{2}{*}{Diff. rec. $\x_{T}$} & & \textbf{88.36 $\pm$ 12.60} & 35.67 $\pm$ 34.12 \\
 & & \checkmark & 88.23 $\pm$ 11.69 & \textbf{35.37 $\pm$ 31.79} \\  \hline
\multirow{4}{*}{Abdominal CT} 
& \multirow{2}{*}{No diff.} & & 87.48 $\pm$ 5.02 & 6.63 $\pm$ 4.03 \\
&  & \checkmark & \textbf{87.59 $\pm$ 5.10} & \textbf{6.36 $\pm$ 3.86} \\ \cline{2-5}
 & \multirow{2}{*}{Diff. rec. $\x_{T}$} & & 86.89 $\pm$ 5.49 & 6.91 $\pm$ 4.35 \\ 
 &  & \checkmark & \textbf{\underline{87.45 $\pm$ 5.43}} & \textbf{6.56 $\pm$ 5.44} \\ \hline
\multirow{4}{*}{Prostate MR} 
& \multirow{2}{*}{No diff.} & & 84.82 $\pm$ 5.69 & \textbf{4.55 $\pm$ 2.17} \\
&  & \checkmark & \textbf{\underline{85.22 $\pm$ 5.18}} & 4.62 $\pm$ 2.37 \\ \cline{2-5}
 & \multirow{2}{*}{Diff. rec. $\x_{T}$} & & \textbf{85.63 $\pm$ 5.19} & 4.59 $\pm$ 2.71 \\
 &  & \checkmark & 85.54 $\pm$ 5.20 & \textbf{4.40 $\pm$ 1.96} \\ \hline
\multirow{4}{*}{Brain MR} 
& \multirow{2}{*}{No diff.} & & 92.03 $\pm$ 9.67 & 5.29 $\pm$ 8.53 \\
& & \checkmark & \textbf{\underline{92.43 $\pm$ 9.10}} & \textbf{5.20 $\pm$ 9.56} \\ \cline{2-5}
 & \multirow{2}{*}{Diff. rec. $\x_{T}$} & & 92.04 $\pm$ 9.47 & 7.25 $\pm$ 13.76 \\
& & \checkmark & \textbf{92.29 $\pm$ 8.55} & \textbf{7.03 $\pm$ 13.48} \\
\bottomrule
\end{tabular}
\end{table}
Compared to~\citet{fu2023importance}, the model includes a Transformer layer at the bottom encoder of U-net. This component has one layer representing 16\% and 6\% of the trainable parameters for 2D and 3D networks, correspondingly (see \Cref{tab:network-size} in \Cref{app:implementation-details}). An ablation study has been performed for the proposed recycling approach and non-diffusion models. The results have been summarised in \Cref{tab:transformer_ablation}. For non-diffusion models, the addition of the Transformer component brought improvement in Dice score across all applications ($p<0.001$ for muscle ultrasound; $p>=0.05$ for abdominal CT; $p=0.001$ for prostate MR; and $p=0.0178$ for brain MR), making this architecture the stronger reference model. For diffusion, significantly higher Dice scores have been observed for abdominal CT data ($p<0.001$), and the differences were not significant for other applications ($p>=0.05$).

\subsubsection{Length of training noise schedule}
It's worth noting that~\citet{fu2023importance} recommended incorporating a shortened variance schedule during training, mirroring that used during inference, in addition to the recycling technique. This modification resulted in enhanced performance for every training strategy on the muscle ultrasound data set (as detailed in~\Cref{tab:timesteps_ablation_muscle}). However, this adaptation did not yield enhancements for the proposed training strategies (``Diff. rec. $\x_T$'') in the abdominal CT data set (as depicted in~\Cref{tab:timesteps_ablation_abdominal}). Moreover, not all differences observed were statistically significant. This may suggest that the advantage of the modified training variance schedule may be application-dependent and sensitive to the change of model architectures and hyper-parameters. In this work, the variance schedule was maintained at $1001$ steps.

\section{Conclusion}
In this research, we have proposed a novel training strategy for diffusion-based segmentation models. The aim is to remove the dependency on ground truth masks during denoising training. In contrast to the standard diffusion-based segmentation models and those employing self-conditioning or alternative recycling techniques, our approach consistently maintains or enhances segmentation performance throughout progressive inference processes. Through extensive experiments across four medical imaging data sets with different dimensionalities and modalities, we demonstrated statistically significant improvement against all diffusion baseline models for both DDPM and DDIM samplers. Our analysis for the first time identified a common limitation of existing diffusion model training for segmentation tasks. The use of ground truth data for denoising training leads to data leakage. By utilizing the model's prediction at the initial step instead, we align the training process with inference procedures, effectively reducing over-fitting and promoting better generalization. While existing diffusion models underperformed non-diffusion-based segmentation model baselines, our innovative recycling training strategies effectively bridged the performance gap. This enhancement allowed diffusion models to attain comparable performance levels. To the best of our knowledge, this is the first time diffusion models have achieved such parity in performance while maintaining identical architecture and compute budget. By ensembling the diffusion and non-diffusion models, constant and significant improvements have been observed across all data sets, demonstrating one of its potential values. Nevertheless, challenges remain on the road to advancing diffusion-based segmentation models further. Future work could explore discrete diffusion models that are tailored for categorical data or implement diffusion in latent space to further reduce compute costs. Although the presented experimental results primarily demonstrated methodological development, the fact that these were obtained on four large clinical data sets represents a promising step toward real-world applications. We would like to argue the potential importance of the reported development, which may lead to better clinical outcomes and improved patient care in respective applications. For example, avoiding surrounding healthy structures may be sensitive to their localization in planning imaging, in both the abdominal CT and prostate MR tasks. This sensitivity can be high and nonlinear therefore arguably a perceived marginal improvement might benefit those with smaller targets, such as those in liver resection and focal therapy of prostate cancer, or highly variable ultrasound imaging guidance.


\acks{This work was supported by the EPSRC grant (EP/T029404/1), the Wellcome/EPSRC Centre for Interventional and Surgical Sciences (203145Z/16/Z), the International Alliance for Cancer Early Detection, an alliance between Cancer Research UK (C28070/A30912, C73666/A31378), Canary Center at Stanford University, the University of Cambridge, OHSU Knight Cancer Institute, University College London and the University of Manchester, and Cloud TPUs from Google's TPU Research Cloud (TRC).}

%
\ethics{The work follows appropriate ethical standards in conducting research and writing the manuscript, following all applicable laws and regulations regarding treatment of animals or human subjects.}

\coi{We declare we do not have conflicts of interest.}

\bibliography{references}


\clearpage
\appendix

\section{Denoising Diffusion Probabilistic Model}
\label{app:ddpm}
We review the formulation of denoising diffusion probabilistic models (DDPM) from~\citet{sohl2015deep,ho2020denoising,nichol2021improved}.
\subsection{Definition}
\begin{align*}
    \x_T \autorightleftharpoons{}{} \cdots \autorightleftharpoons{}{} \x_t \autorightleftharpoons{$p_\theta(\x_{t-1}\mid\x_t)$}{$q(\x_t\mid\x_{t-1})$} \x_{t-1} \autorightleftharpoons{}{} \cdots \autorightleftharpoons{}{} \x_0
\end{align*}
Consider a continuous \textit{diffusion} process (also named \textit{forward} process or \textit{noising} process): given a data point $\x_0\sim q(\x_0)$ in $\R^D$, we add noise to $\x_t$ for $t=1,\cdots,T$ with the following multivariate normal distribution:
\begin{align*}
    q(\x_t\mid\x_{t-1})= \mathcal{N}(\x_t;\sqrt{1-\beta_t}\x_{t-1}, \beta_t\mathbf{I})
\end{align*}
where $\beta_t\in[0,1]$ is a variance schedule. Given sufficiently large $T$ and a well-defined variance schedule, the distribution of $\x_T$ approximates an isotropic multivariate normal distribution.
\begin{align*}
    q(\x_t\mid\x_0) \rightarrow \mathcal{N}(\x_t;\mathbf{0}, \mathbf{I})
\end{align*}
Therefore, we can define a \textit{reverse} process (also named \textit{denoising} process): given a sample $\x_T\sim\mathcal{N}(\x_T;\mathbf{0},\mathbf{I})$, we denoise the data using neural networks $\bm{\mu}_\theta:\R^D\rightarrow \R^D$ and $\mathbf{\Sigma}_\theta:\R^D\rightarrow \R^{D\times D}$ as follows:
\begin{align*}
    p_\theta(\x_{t-1}\mid\x_{t})= \mathcal{N}(\x_{t-1};\bm{\mu}_\theta(\x_t,t),\mathbf{\Sigma}_\theta(\x_t,t))
\end{align*}
In this work, an isotropic variance is assumed with $\mathbf{\Sigma}_\theta(\x_t,t)=\sigma_t^2\mathbf{I}$, such that
\begin{align*}
    p_\theta(\x_{t-1}\mid\x_{t})= \mathcal{N}(\x_{t-1};\bm{\mu}_\theta(\x_t,t),\sigma_t^2\mathbf{I})
\end{align*}

\subsection{Variational Lower Bound}
Consider $\z=\x_{1:T}\mid\x_0$ as latent variables for $\x_0$, we can derive the variational lower bound (VLB) as follows:
\begin{align*}
\log p_\theta(\x_0)
=& D_\text{KL}(q(\z)~\|~p_\theta(\z\mid\x_0)) + \E_{q(\z)} \left[\log \frac{p_\theta(\x_0, \z)}{q(\z)}\right]\\
\geq& \E_{q(\z)} \left[\log \frac{p_\theta(\x_0, \z)}{q(\z)}\right]\\
=&-\Big(\E_{q(\x_1\mid\x_0)}L_0 +\sum_{t=2}^T \E_{q(\x_t\mid\x_{0})}L_{t-1} + L_T\Big)
\end{align*}
where
\begin{align*}
    L_0 &= - \log p_\theta(\x_0\mid\x_1)&\text{(reconstruction loss)} \\
    L_{t-1} &= D_\text{KL}(q(\x_{t-1}\mid\x_{t},\x_{0})\|p_\theta(\x_{t-1}\mid\x_{t}))&\text{(diffussion loss)} \\
    L_T &= D_\text{KL}(q(\x_T\mid\x_{0}))~\|~p_\theta(\x_T).&\text{(prior loss)}
\end{align*}

\subsection{Diffusion Loss}
\label{subsec:ddpm-diffusion-loss}
In particular, we can derive the closed form $L_{t-1}$ with
\begin{align*}
    q(\x_t\mid\x_0) &= \mathcal{N}(\x_t;\sqrt{\bar{\alpha}_t}\x_0, (1-\bar{\alpha}_t)\mathbf{I}) \\
    q(\x_{t-1}\mid\x_t, \x_0) &= \mathcal{N}(\x_{t-1};\tilde{\bm{\mu}}(\x_t,\x_0),\tilde{\beta}_t\mathbf{I})
\end{align*}
where
\begin{align}
    \alpha_t &= 1-\beta_t \nonumber \\
    \bar{\alpha}_t &= \prod_{s=1}^t \alpha_s \nonumber \\
    \tilde{\bm{\mu}}(\x_t,\x_0)&=\frac{\sqrt{\bar{\alpha}_{t-1}}\beta_t}{1-\bar{\alpha}_{t}}\x_{0}
    +\frac{1 - \bar{\alpha}_{t-1}}{1-\bar{\alpha}_{t}}\sqrt{\alpha_t}\x_{t},\label{eq:ddpm-diff-loss-mu}\\
    \tilde{\beta}_t&=\frac{1 - \bar{\alpha}_{t-1}}{1-\bar{\alpha}_{t}}\beta_t. \nonumber
\end{align}

\subsubsection{Noise Prediction Loss ($\epsilon$-parameterization)s}
\label{subsubsec:ddpm-noise-prediction-loss}
Consider the reparameterization in~\citet{ho2020denoising},
\begin{align*}
    \x_t(\x_0,\eps)&=\sqrt{\bar{\alpha}_t}\x_0+\sqrt{1-\bar{\alpha}_t}\eps\\
        \eps_\theta(\x_t,t) &= 
\frac{1}{\sqrt{1-\bar{\alpha}_t}}\x_t - \frac{\sqrt{\bar{\alpha}_t}}{\sqrt{1-\bar{\alpha}_t}}\x_0\\
    \bm{\mu}_\theta(\x_t,t)&=\frac{1}{\sqrt{\alpha}_t}(\x_{t}
    -\frac{\beta_t}{\sqrt{1-\bar{\alpha}_t}}\eps_\theta(\x_t,t))
\end{align*}
We can derive a closed form of $L_{t-1}$
\begin{align*}
    L_{t-1}(\x_{t},\x_{0})
    =\frac{1}{2\sigma_t^2}\frac{\beta_t^2}{{\alpha_t(1-\bar{\alpha}_t)}}\|\eps
    -\eps_\theta\|_2^2 + C
\end{align*}

If $\sigma_t^2 = \tilde{\beta}_t = \frac{1 - \bar{\alpha}_{t-1}}{1-\bar{\alpha}_{t}}\beta_t$, using the signal-to-noise ratio (SNR) defined in ~\citet{kingma2021variational}, $\text{SNR}(t) = \frac{\bar{\alpha}_t}{1-\bar{\alpha}_t}$, the loss can be derived as
\begin{align*}
    L_{t-1}(\x_{t},\x_{0})
    =(\frac{\text{SNR}(t-1)}{\text{SNR}(t)}-1)\|\eps
    -\eps_\theta\|_2^2 + C
\end{align*}

\subsubsection{Sample Prediction Loss ($\x_0$-parameterization)}
Similar to \cref{eq:ddpm-diff-loss-mu}, consider the parameterization~\citep{kingma2021variational},
\begin{align*}
    \bm{\mu}_\theta(\x_t,t)=\frac{\sqrt{\bar{\alpha}_{t-1}}\beta_t}{1-\bar{\alpha}_{t}}\x_{0,\theta}
    +\frac{1 - \bar{\alpha}_{t-1}}{1-\bar{\alpha}_{t}}\sqrt{\alpha_t}\x_{t}
\end{align*}
We can derive a closed form of $L_{t-1}$
\begin{align*}
    L_{t-1}(\x_{t},\x_{0})=\frac{1}{2\sigma_t^2}\frac{\bar{\alpha}_{t-1}\beta_t^2}{(1-\bar{\alpha}_{t})^2}\|\x_{0,\theta}-\x_{0}\|_2^2 + C.
\end{align*}

If $\sigma_t^2 = \tilde{\beta}_t = \frac{1 - \bar{\alpha}_{t-1}}{1-\bar{\alpha}_{t}}\beta_t$, using the signal-to-noise ratio (SNR) defined in ~\citet{kingma2021variational}, $\text{SNR}(t) = \frac{\bar{\alpha}_t}{1-\bar{\alpha}_t}$, the loss can be derived as
\begin{align*}
    L_{t-1}(\x_{t},\x_{0})=\frac{1}{2}(\text{SNR}(t-1) - \text{SNR}(t))\|\x_{0,\theta}-\x_{0}\|_2^2 + C.
\end{align*}

\subsection{Training}
Empirically, instead of using the variational lower bound, the neural network can be trained on one of the following simplified loss~\citep{ho2020denoising}
\begin{align*}
    L_{\text{simple},\eps_t}(\theta)
    &=\E_{t,\x_0,\eps_t}\|\eps_t
    -\eps_{t,\theta}\|_2^2=\E_{t,\x_0,\eps_t}L(\eps_t
    ,\eps_{t,\theta}),~(\eps\text{-parameterization})\\
    L_{\text{simple},\x_0}(\theta)
    &=\E_{t,\x_0,\eps_t}\|\x_{0}-\x_{0,\theta}\|_2^2=\E_{t,\x_0,\eps_t}L(\x_{0},\x_{0,\theta}).~(\x_0\text{-parameterization})
\end{align*}
with $t$ uniformly sampled from $1$ to $T$ and $\eps_t\sim\mathcal{N}(\mathbf{0},\mathbf{I})$. $L(\cdot,\cdot)$ is a loss function in the space of $\x$. With the importance sampling proposed in~\citet{nichol2021improved}, $t$ can be sampled with a probability proportional to $\E_{\x_0,\x_t}L(\x_{0},\hat{\x}_0)$. In other words, a time step $t$ is sampled more often if the loss is larger.

As the previous work~\citep{fu2023importance} has extensively compared the $\eps$-parameterization and $\x_0$-parameterization, as well as the benefits of including Dice loss, in this work, we use $\x_0$-parameterization with a weighted sum of cross-entropy and foreground-only Dice loss~\citet{kirillov2023segment}. 

\subsection{Variance Resampling}
Given a variance schedule $\{\beta_t\}_{t=1}^T$ (e.g. $T=1001$), a subsequence $\{\beta_k\}_{k=1}^K$ (e.g. $K=5$) can be sampled with $\{t_k\}_{k=1}^K$. Following \citet{nichol2021improved}, we can define $\beta_k=1-\frac{\bar{\alpha}_{t_k}}{\bar{\alpha}_{t_{k-1}}}$ then $\alpha_k = 1-\beta_k$ and $\bar{\alpha}_k = \prod_{s=1}^k \alpha_s$ can be recalculated correspondingly. In this work, $t_k$ is uniformly downsampled. For instance, if $T=1001$ and $K=5$, then $\{t_k\}_{k=1}^K=\{1,251,501,751,1001\}$.

\section{Denoising Diffusion Implicit Model}
\label{app:ddim}
\paragraph{Definition}~\citet{song2020denoising} parameterize $q(\x_{t-1}\mid\x_t, \x_0)$ as follows, with $\eps=\frac{\x_t-\sqrt{\bar{\alpha}_t}\x_0}{\sqrt{1-\bar{\alpha}_t}}$,
\begin{align*}
    q(\x_{t-1}\mid\x_t, \x_0) 
    &= \mathcal{N}(\x_{t-1};\sqrt{\bar{\alpha}_{t-1}}\x_0
    +\sqrt{1 - \bar{\alpha}_{t-1}-\sigma_t}\eps,\sigma_t^2\mathbf{I}).
\end{align*}
For any variance schedule $\sigma_t$, this formulation ensures $q(\x_t\mid\x_0)=\mathcal{N}(\x_t;\sqrt{\bar{\alpha}_t}\x_0, (1-\bar{\alpha}_t)\mathbf{I})$. Particularly, if $\sigma_t^2=\tilde{\beta}_t$, this represents DDPM. If $\sigma_t=0$ for $t>1$ and $\sigma_1=\sqrt{\tilde{\beta}_1}$, the model is deterministic and named as denoising diffusion implicit model (DDIM).

\paragraph{Inference} For DDIM, at inference time, the denoising starts with a Gaussian noise $\x_T\sim\mathcal{N}(\bm{0},\bm{I})$ and the data is denoised step-by-step for $t=T,\cdots,1$:
\begin{align*}
p_\theta(\x_{t-1}\mid\x_{t})=\begin{cases}
\mathcal{N}(\hat{\x}_0, \sigma_1^2\mathbf{I}) & t = 1\\
q(\x_{t-1}\mid\x_t, \x_{0,\theta}(\x_t,t)) & t>1
\end{cases}
\end{align*}

\section{Self-conditioning}
\label{app:self-conditioning}
The self-conditioning methods proposed in~\citet{chen2022analog} (``Diff. sc. $\x_t$'' in~\Cref{eq:ddpm-train-segmentation-self-conditioning-same}) and~\citet{watson2023novo} (``sc. $\x_{t+1}$'' in~\Cref{eq:ddpm-train-segmentation-self-conditioning-next}) are illustrated below.
\begin{subequations}\label{eq:ddpm-train-segmentation-self-conditioning-same}
\begin{align}
    \x_t &\sim \mathcal{N}(\x_t;\sqrt{\bar{\alpha}_t}\mathbf{x}_0, (1-\bar{\alpha}_t)\mathbf{I}),&(\text{sc.}~\x_{t},~\text{step 1, sampling}) \label{eq:sc-same-step1-sample}\\
    \hat{\mathbf{x}}_0 &= \text{StopGradient}(f_\theta(I,t,\mathbf{x}_{t},\mathbf{0})),&(\text{sc.}~\x_{t},~\text{step 1, prediction}) \label{eq:sc-same-step1-pred}\\
    \hat{\mathbf{x}}_0 &= \text{Dropout}_{p=50\%}(\hat{\mathbf{x}}_0),&(\text{sc.}~\x_{t},~\text{step 2, dropout})  \label{eq:sc-same-step2-drop}\\
    \hat{\mathbf{x}}_0 &= f_\theta(I,t,\mathbf{x}_t, \hat{\mathbf{x}}_0),&(\text{sc.}~\x_{t},~\text{step 2, prediction}) \label{eq:sc-same-step2-pred}\\
    L_\text{denoising}(\theta)
    &=\E_{t,\x_0,\x_t}L(\x_{0},\hat{\mathbf{x}}_0),&(\text{loss calculation})
\end{align}
\end{subequations}
\begin{subequations}\label{eq:ddpm-train-segmentation-self-conditioning-next}
\begin{align}
    \x_{t+1} &\sim \mathcal{N}(\x_{t+1};\sqrt{\bar{\alpha}_{t+1}}\x_0, (1-\bar{\alpha}_{t+1})\mathbf{I}),&(\text{sc.}~\x_{t+1},~\text{step 1, sampling}) \label{eq:sc-next-step1-sample}\\
    \hat{\mathbf{x}}_0 &= \text{StopGradient}(f_\theta(I,t+1,\mathbf{x}_{t+1},\mathbf{0})),&(\text{sc.}~\x_{t+1},~\text{step 1, prediction}) \label{eq:sc-next-step1-pred}\\
    \hat{\mathbf{x}}_0 &= \text{Dropout}_{p=50\%}(\hat{\mathbf{x}}_0),&(\text{sc.}~\x_{t+1},~\text{step 2, dropout}) \label{eq:sc-next-step2-drop}\\
    \x_t &\sim \mathcal{N}(\x_{t};\tilde{\bm{\mu}},\tilde{\beta}_{t+1}\mathbf{I}),&(\text{sc.}~\x_{t+1},~\text{step 2, sampling}) \label{eq:sc-next-step2-sample}\\
    \tilde{\bm{\mu}}&=\frac{\sqrt{\bar{\alpha}_{t}}\beta_{t+1}}{1-\bar{\alpha}_{t+1}}\x_{0}
    +\frac{1 - \bar{\alpha}_{t}}{1-\bar{\alpha}_{t+1}}\sqrt{\alpha_{t+1}}\x_{t+1} \nonumber\\
    \hat{\mathbf{x}}_0 &= f_\theta(I,t,\mathbf{x}_t, \hat{\mathbf{x}}_0),&(\text{sc.}~\x_{t+1},~\text{step 2, prediction}) \label{eq:sc-next-step2-pred}\\
    L_\text{denoising}(\theta)
    &=\E_{t,\x_0,\x_t}L(\x_{0},\hat{\mathbf{x}}_0),&(\text{loss calculation})
\end{align}
\end{subequations}

\section{Diffusion Noise Schedule}
\label{app:noise}
\begin{figure}[!h]
     \centering
     \begin{subfigure}[b]{0.4\textwidth}
         \centering
         \includegraphics[width=\textwidth]{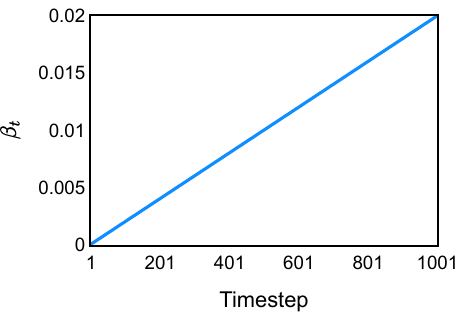}
         \caption{$\beta_t$}
     \end{subfigure}
     \begin{subfigure}[b]{0.4\textwidth}
         \centering
         \includegraphics[width=\textwidth]{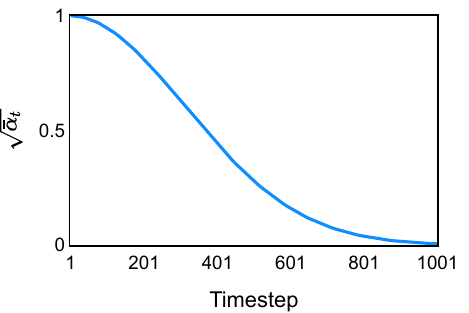}
         \caption{$\sqrt{\bar{\alpha_t}}$}
     \end{subfigure}
     \begin{subfigure}[b]{0.4\textwidth}
         \centering
         \includegraphics[width=\textwidth]{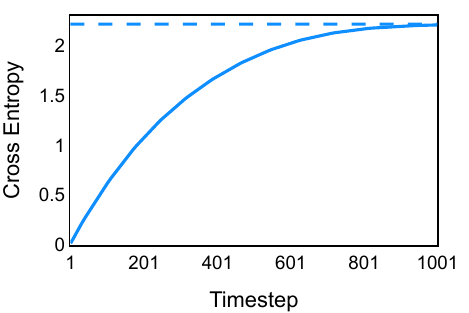}
         \caption{Cross entropy}
     \end{subfigure}
     \begin{subfigure}[b]{0.4\textwidth}
         \centering
         \includegraphics[width=\textwidth]{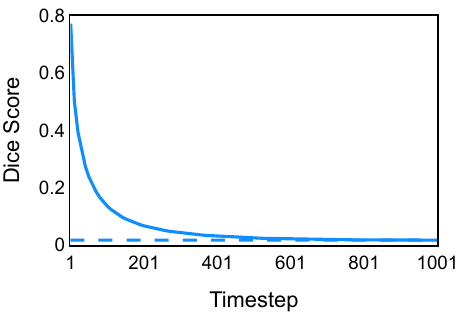}
         \caption{Dice score}
     \end{subfigure}
    \caption{\textbf{Information contained in $\x_t$.} Cross entropy and dice score between $\x_t$ and ground truth $\x_0$ are used to empirically measure the amount of information of ground truth $\x_0$ contained in $\x_t$. The dashed line represents the information contained in the sampled noise (between noise and ground truth $\x_0$), which is considered to be the limit. The values are calculated using the sample ``005095'' in prostate MR data set.}
    \label{fig:x_t}
\end{figure}

The noise schedule $\beta_t$ and $\sqrt{\bar{\alpha}_t}$ have been visualised in \Cref{fig:x_t}. The cross entropy and dice score between $\x_t$ and ground truth $\x_0$ have also been visualized to empirically measure the amount of information of ground truth $\x_0$ contained in $\x_t$.

\section{Implementation Details}
\label{app:implementation-details}
\begin{table}[!ht]
    \centering
    \caption{\textbf{Training Hyper-parameters}}
    \label{tab:training-hparams}
    \begin{tabular}{c|c}
        \toprule
        Parameter & Value \\
        \midrule
        Optimiser & AdamW (b1=0.9, b2=0.999, weight\_decay=1E-8) \\
        Learning Rate Warmup &  100 steps \\
        Learning Rate Decay &  10,000 steps \\
        Learning Rate Values & Initial = 1E-5, Peak = 8E-4, End = 5E-5 \\
        Batch size & 256 for Muscle Ultrasound and 8 for other data sets \\
        Number of samples & 320K for Muscle Ultrasound and 100K for other data sets \\
        \bottomrule
    \end{tabular}
\end{table}

\begin{table}[!ht]
\centering
\caption{\textbf{Network Size}}
\label{tab:network-size}
\begin{tabular}{l|l|c|c}
\toprule
\multirow{2}{*}{Dimension} & \multirow{2}{*}{Method} & \multicolumn{2}{c}{Transformer} \\ \cline{3-4}
 & & \checkmark              &                \\
\midrule
\multirow{2}{*}{2D}     & No diff.    & 12,586,594     & 10,550,370     \\ \cline{2-4}
                        & Diff.       & 13,335,554     & 11,299,330     \\ \hline
\multirow{2}{*}{3D}     & No diff.    & 33,385,154     & 31,283,394     \\ \cline{2-4}
                        & Diff.       & 34,135,266     & 32,033,506    \\
\bottomrule
\end{tabular}
\end{table}
\clearpage
\newpage
\section{Results}
\subsection{Diffusion Training Strategy Comparison}
\label{app:results-diff}
\begin{table}[!ht]
\centering
\caption{\textbf{Per class Dice score comparison.} ``No diff.'' represents non-diffusion model. ``Diff. rec. $\x_T$'' represents the diffusion model with proposed recycling. ``Ensemble'' represents the model averaging the probabilities from ``No diff.'' and ``Diff. rec. $\x_T$''. The inference sampler is DDPM. The best results are in bold and underline indicates the difference to non-diffusion model is significant with p-value < $0.05$.}
\label{tab:per_class_dice}

\begin{subtable}[!ht]{\textwidth}
\centering
\begin{tabular}{l|cccc}
\toprule
Method & Spleen & RT kidney & LT kidney & Gall bladder \\
\midrule
No diff. & 96.62 $\pm$ 1.87 & 95.08 $\pm$ 10.74 & 96.29 $\pm$ 1.73 & 78.83 $\pm$ 27.82 \\
Diff. rec. $\x_{T}$ & 96.40 $\pm$ 2.42 & 96.24 $\pm$ 1.90 & 96.27 $\pm$ 1.53 & 76.68 $\pm$ 29.25 \\
Ensemble & \textbf{96.78 $\pm$ 1.75} & \textbf{\underline{96.47 $\pm$ 2.44}} & \textbf{\underline{96.50 $\pm$ 1.51}} & \textbf{79.65 $\pm$ 27.29} \\
\midrule
Method & Esophagus & Liver & Stomach & Arota \\
\midrule
No diff. & 83.22 $\pm$ 11.08 & 97.36 $\pm$ 1.17 & 90.53 $\pm$ 14.78 & 94.65 $\pm$ 4.22 \\
Diff. rec. $\x_{T}$ & 83.60 $\pm$ 10.32 & 97.33 $\pm$ 1.13 & 90.77 $\pm$ 14.46 & 94.66 $\pm$ 4.66 \\
Ensemble & \textbf{\underline{84.10 $\pm$ 11.15}} & \textbf{\underline{97.54 $\pm$ 1.05}} & \textbf{\underline{91.07 $\pm$ 14.91}} & \textbf{\underline{94.96 $\pm$ 4.39}} \\
\midrule
Method & Postcava & Pancreas & Right adrenal gland & Left adrenal gland \\
\midrule
No diff. & 90.45 $\pm$ 4.68 & 84.88 $\pm$ 11.40 & 77.80 $\pm$ 9.46 & 77.98 $\pm$ 11.95 \\
Diff. rec. $\x_{T}$ & 90.55 $\pm$ 4.19 & 84.86 $\pm$ 11.15 & \underline{76.63 $\pm$ 12.84} & 78.01 $\pm$ 11.60 \\
Ensemble & \textbf{\underline{91.18 $\pm$ 4.12}} & \textbf{\underline{85.85 $\pm$ 11.12}} & \textbf{\underline{78.51 $\pm$ 10.58}} & \textbf{\underline{78.95 $\pm$ 11.45}} \\
\midrule
Method & Duodenum & Bladder & Prostate/uterus \\
\midrule
No diff. & 79.57 $\pm$ 14.89 & 88.09 $\pm$ 16.25 & 82.35 $\pm$ 18.90 \\
Diff. rec. $\x_{T}$ & 79.80 $\pm$ 15.14 & 87.90 $\pm$ 16.65 & 81.90 $\pm$ 18.86 \\
Ensemble & \textbf{\underline{80.99 $\pm$ 15.07}} & \textbf{\underline{88.61 $\pm$ 16.59}} & \textbf{\underline{83.06 $\pm$ 18.68}} \\
\bottomrule
\end{tabular}
\caption{Abdominal CT:  LT and RT stand for left and right, respectively.}
\label{tab:abdominal_ct_per_class_dice}
\end{subtable}

\begin{subtable}[!ht]{\textwidth}
\centering
\begin{tabular}{l|cccc}
\toprule
Method & Bladder & Bone & Obturator internus & Transition zone \\
\midrule
No diff. & 93.28 $\pm$ 9.90 & 93.12 $\pm$ 5.68 & 88.95 $\pm$ 3.53 & 79.61 $\pm$ 8.37 \\
Diff. rec. $\x_{T}$ & 93.57 $\pm$ 9.61 & \textbf{\underline{93.84 $\pm$ 5.85}} & \underline{89.15 $\pm$ 3.62} & 79.79 $\pm$ 8.36 \\
Ensemble & \textbf{\underline{93.66 $\pm$ 9.84}} & \underline{93.77 $\pm$ 5.52} & \textbf{\underline{89.52 $\pm$ 3.51}} & \textbf{\underline{80.57 $\pm$ 8.20}} \\
\midrule
Method & Central gland & Rectum & Seminal vesicle & NV bundle \\
\midrule
No diff. & 88.75 $\pm$ 5.60 & 93.30 $\pm$ 3.48 & 77.55 $\pm$ 10.99 & 67.17 $\pm$ 14.34 \\
Diff. rec. $\x_{T}$ & \underline{89.13 $\pm$ 5.78} & 93.42 $\pm$ 3.51 & \underline{78.39 $\pm$ 9.71} & 67.07 $\pm$ 15.50 \\
Ensemble & \textbf{\underline{89.45 $\pm$ 5.56}} & \textbf{\underline{93.70 $\pm$ 3.37}} & \textbf{\underline{78.91 $\pm$ 10.28}} & \textbf{\underline{68.01 $\pm$ 14.85}} \\
\bottomrule
\end{tabular}
\caption{Prostate MR: Dice score per class. NV stands for neurovascular.}
\label{tab:prostate_mr_per_class_dice}
\end{subtable}
\end{table}

\begin{table}[!ht]
\centering
\caption{\textbf{Per class Hausdorff distance comparison} ``No diff.'' represents non-diffusion model. ``Diff. rec. $\x_T$'' represents the diffusion model with proposed recycling. ``Ensemble'' represents the model averaging the probabilities from ``No diff.'' and ``Diff. rec. $\x_T$''. The inference sampler is DDPM. The best results are in bold and underline indicates the difference to non-diffusion model is significant with p-value < $0.05$.}
\label{tab:per_class_hausdorff}

\begin{subtable}[!ht]{\textwidth}
\centering
\begin{tabular}{l|cccc}
\toprule
Method & Spleen & Right kidney & Left kidney & Gall bladder \\
\midrule
No diff. & 3.22 $\pm$ 4.91 & 1.97 $\pm$ 1.46 & 4.13 $\pm$ 10.83 & \textbf{9.23 $\pm$ 16.71} \\
Diff. rec. $\x_{T}$ & \textbf{2.86 $\pm$ 3.84} & 1.93 $\pm$ 0.83 & 3.13 $\pm$ 8.35 & 12.65 $\pm$ 21.74 \\
Ensemble & \underline{2.89 $\pm$ 4.28} & \textbf{\underline{1.84 $\pm$ 1.11}} & \textbf{2.70 $\pm$ 5.86} & 9.57 $\pm$ 18.86 \\
\midrule
Method & Esophagus & Liver & Stomach & Arota \\
\midrule
No diff. & 5.50 $\pm$ 6.81 & 3.50 $\pm$ 2.50 & 8.96 $\pm$ 13.99 & 6.62 $\pm$ 14.52 \\
Diff. rec. $\x_{T}$ & 5.30 $\pm$ 6.41 & 3.79 $\pm$ 4.16 & 9.00 $\pm$ 14.03 & \textbf{5.41 $\pm$ 11.20} \\
Ensemble & \textbf{5.22 $\pm$ 6.63} & \textbf{\underline{3.06 $\pm$ 1.63}} & \textbf{8.04 $\pm$ 12.87} & 5.47 $\pm$ 11.29 \\
\midrule
Method & Postcava & Pancreas & Right adrenal gland & Left adrenal gland \\
\midrule
No diff. & 4.80 $\pm$ 4.55 & 7.57 $\pm$ 8.62 & \textbf{4.39 $\pm$ 2.39} & 5.15 $\pm$ 5.40 \\
Diff. rec. $\x_{T}$ & 4.62 $\pm$ 3.09 & 7.50 $\pm$ 8.62 & 4.66 $\pm$ 3.14 & 4.87 $\pm$ 4.64 \\
Ensemble & \textbf{4.41 $\pm$ 3.25} & \textbf{\underline{6.96 $\pm$ 8.40}} & 4.41 $\pm$ 2.79 & \textbf{\underline{4.82 $\pm$ 4.92}} \\
\midrule
Method & Duodenum & Bladder & Prostate/uterus \\
\midrule
No diff. & 10.54 $\pm$ 8.44 & 9.10 $\pm$ 23.07 & 10.97 $\pm$ 19.01 \\
Diff. rec. $\x_{T}$ & \underline{9.31 $\pm$ 7.13} & 10.70 $\pm$ 31.83 & 13.35 $\pm$ 32.75 \\
Ensemble & \textbf{\underline{9.29 $\pm$ 7.37}} & \textbf{6.52 $\pm$ 10.34} & \textbf{9.14 $\pm$ 13.11} \\
\bottomrule
\end{tabular}
\caption{Abdominal CT:  LT and RT stand for left and right, respectively.}
\label{tab:abdominal_ct_per_class_hausdorff}
\end{subtable}

\begin{subtable}[!ht]{\textwidth}
\centering
\begin{tabular}{l|cccc}
\toprule
Method & Bladder & Bone & Obturator internus & Transition zone \\
\midrule
No diff. & 3.30 $\pm$ 4.54 & 3.18 $\pm$ 9.77 & 4.60 $\pm$ 3.36 & 5.97 $\pm$ 4.97 \\
Diff. rec. $\x_{T}$ & 3.20 $\pm$ 4.12 & \textbf{2.21 $\pm$ 1.62} & 4.50 $\pm$ 3.46 & 6.25 $\pm$ 4.96 \\
Ensemble & \textbf{2.95 $\pm$ 3.48} & 2.32 $\pm$ 1.46 & \textbf{\underline{4.34 $\pm$ 3.29}} & \textbf{6.18 $\pm$ 5.11} \\
\midrule
Method & Central gland & Rectum & Seminal vesicle & NV bundle \\
\midrule
No diff. & 3.94 $\pm$ 2.28 & 4.46 $\pm$ 5.69 & 4.82 $\pm$ 3.85 & 6.68 $\pm$ 6.33 \\
Diff. rec. $\x_{T}$ & \underline{3.70 $\pm$ 1.93} & 4.25 $\pm$ 4.75 & 4.57 $\pm$ 2.66 & 6.55 $\pm$ 6.28 \\
Ensemble & \textbf{\underline{3.66 $\pm$ 1.93}} & \textbf{\underline{4.16 $\pm$ 5.25}} & \textbf{4.52 $\pm$ 2.83} & \textbf{\underline{6.45 $\pm$ 6.34}} \\
\bottomrule
\end{tabular}
\caption{Prostate MR: Dice score per class. NV stands for neurovascular.}
\label{tab:prostate_mr_per_class_hausdorff}
\end{subtable}
\end{table}

\begin{figure}[!ht]
     \centering
     \begin{subfigure}[b]{0.45\textwidth}
         \centering
         \includegraphics[width=\textwidth]{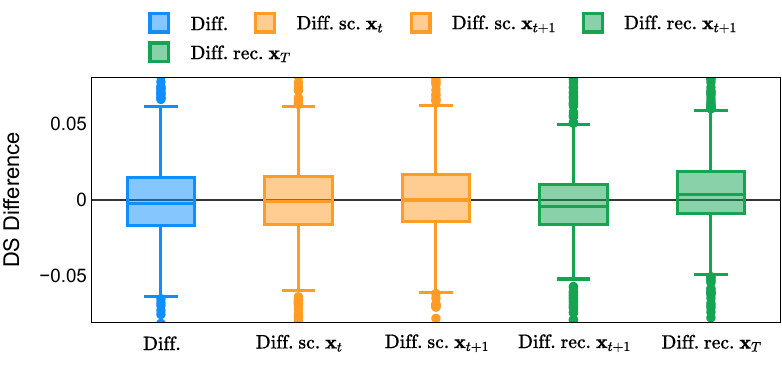}
         \caption{DS difference for muscle ultrasound}
         \label{fig:first_last_step_diff_muscle_us_DDPM_DS}
     \end{subfigure}
     \begin{subfigure}[b]{0.45\textwidth}
         \centering
         \includegraphics[width=\textwidth]{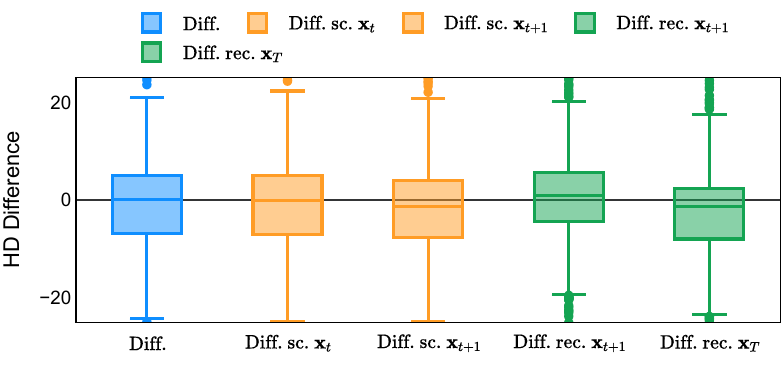}
         \caption{HD difference for muscle ultrasound}
         \label{fig:first_last_step_diff_muscle_us_DDPM_HD}
     \end{subfigure}
     \begin{subfigure}[b]{0.45\textwidth}
         \centering
         \includegraphics[width=\textwidth]{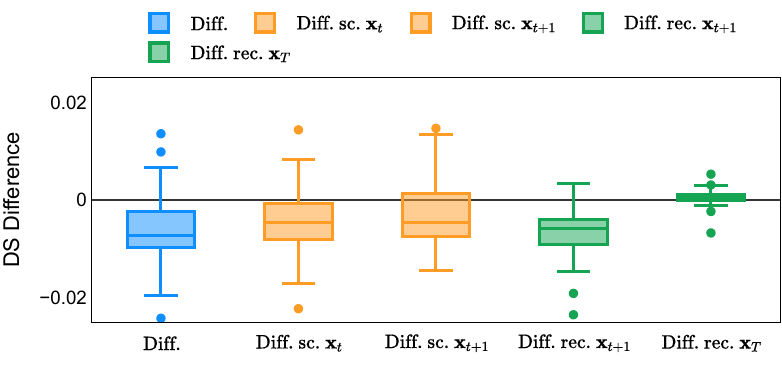}
         \caption{DS difference for abdominal CT}
         \label{fig:first_last_step_diff_amos_ct_DDPM_DS}
     \end{subfigure}
     \begin{subfigure}[b]{0.45\textwidth}
         \centering
         \includegraphics[width=\textwidth]{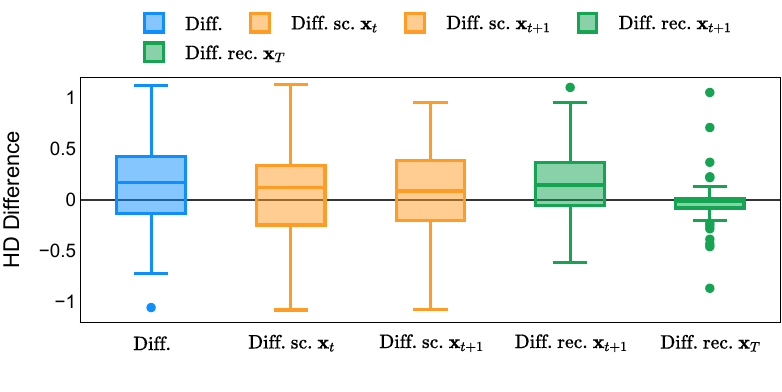}
         \caption{HD difference for abdominal CT}
         \label{fig:first_last_step_diff_amos_ct_DDPM_HD}
     \end{subfigure}
     \begin{subfigure}[b]{0.45\textwidth}
         \centering
         \includegraphics[width=\textwidth]{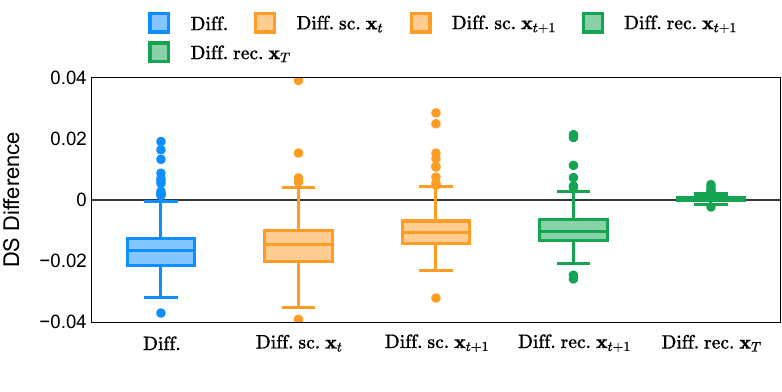}
         \caption{DS difference for prostate MR}
         \label{fig:first_last_step_diff_male_pelvic_mr_DDPM_DS}
     \end{subfigure}
     \begin{subfigure}[b]{0.45\textwidth}
         \centering
         \includegraphics[width=\textwidth]{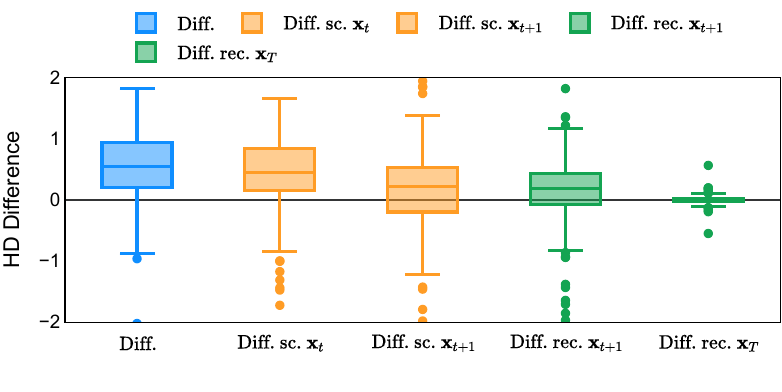}
         \caption{HD difference for prostate MR}
         \label{fig:first_last_step_diff_male_pelvic_mr_DDPM_HD}
     \end{subfigure}
     \begin{subfigure}[b]{0.45\textwidth}
         \centering
         \includegraphics[width=\textwidth]{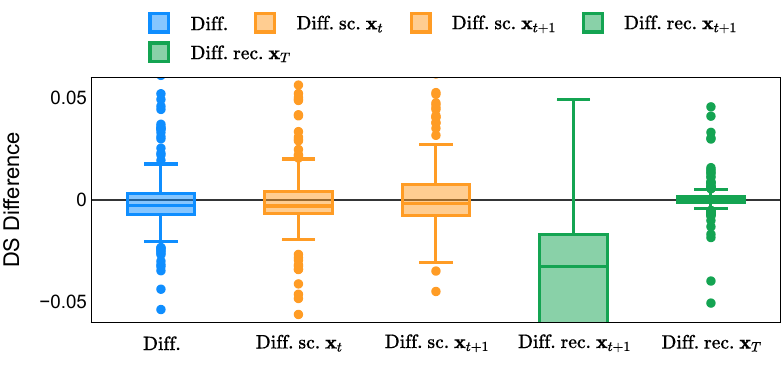}
         \caption{DS difference for brain MR}
         \label{fig:first_last_step_diff_brats2021_mr_DDPM_DS}
     \end{subfigure}
     \begin{subfigure}[b]{0.45\textwidth}
         \centering
         \includegraphics[width=\textwidth]{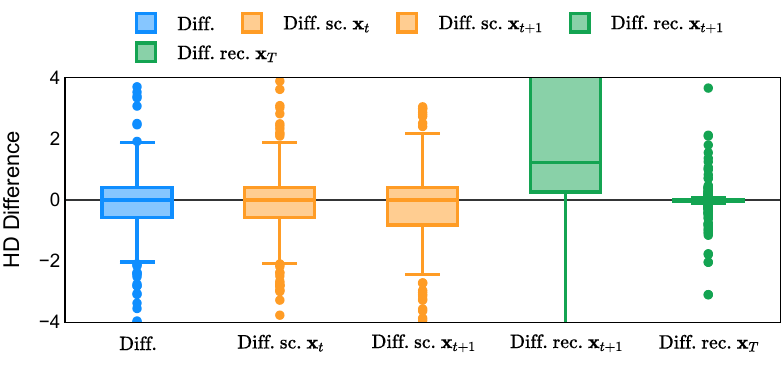}
         \caption{HD difference for brain MR}
         \label{fig:first_last_step_diff_brats2021_mr_DDPM_HD}
     \end{subfigure}
    \caption{\textbf{Segmentation performance difference between the last step and first step using DDPM.} ``Diff.'' represents standard diffusion. ``Diff. sc. $\x_{t}$'' and ``Diff. sc. $\x_{t+1}$'' represents self-conditioning from~\citet{chen2022analog} and~\citet{watson2023novo}, respectively. ``Diff. rec. $\x_{t+1}$'' and ``Diff. rec. $\x_T$'' represents recycling from~\citet{fu2023importance} and the proposed recycling in this work, respectively. The sampler is DDPM. DS and HD represents Dice score and Hausdorff distance, respectively. The difference is the value at the last step subtracted by the one at the first step. A positive value for Dice score difference or a negative value for Hausdorff distance means improvement.}
    \label{fig:first_last_step_diff_ddpm}
\end{figure}
\begin{figure}[!ht]
     \centering
     \begin{subfigure}[b]{0.45\textwidth}
         \centering
         \includegraphics[width=\textwidth]{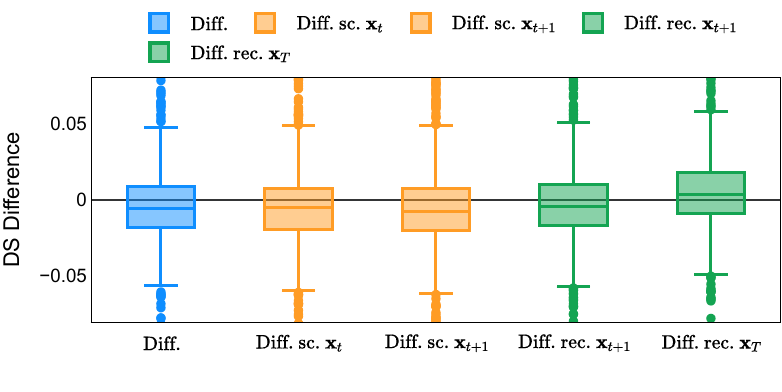}
         \caption{DS difference for muscle ultrasound}
         \label{fig:first_last_step_diff_muscle_us_DDIM_DS}
     \end{subfigure}
     \begin{subfigure}[b]{0.45\textwidth}
         \centering
         \includegraphics[width=\textwidth]{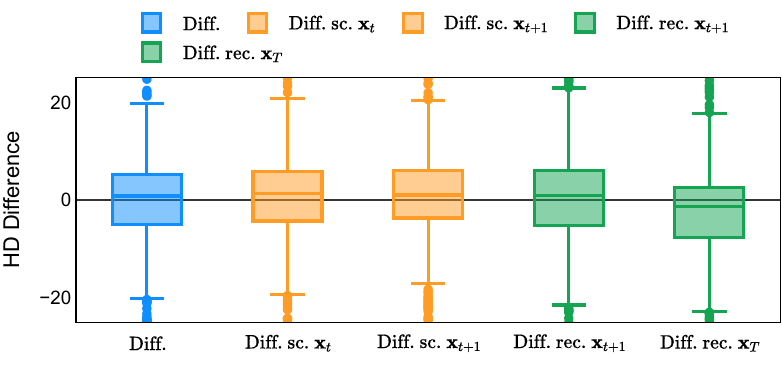}
         \caption{HD difference for muscle ultrasound}
         \label{fig:first_last_step_diff_muscle_us_DDIM_HD}
     \end{subfigure}
     \begin{subfigure}[b]{0.45\textwidth}
         \centering
         \includegraphics[width=\textwidth]{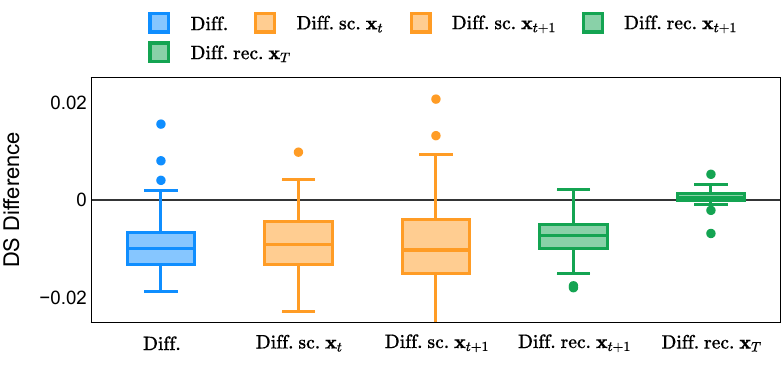}
         \caption{DS difference for abdominal CT}
         \label{fig:first_last_step_diff_amos_ct_DDIM_DS}
     \end{subfigure}
     \begin{subfigure}[b]{0.45\textwidth}
         \centering
         \includegraphics[width=\textwidth]{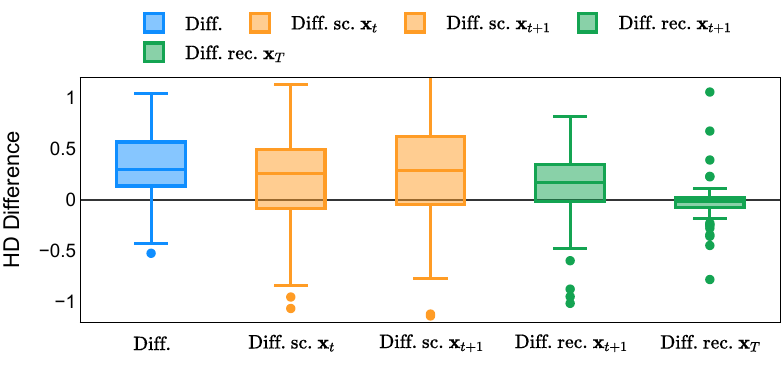}
         \caption{HD difference for abdominal CT}
         \label{fig:first_last_step_diff_amos_ct_DDIM_HD}
     \end{subfigure}
     \begin{subfigure}[b]{0.45\textwidth}
         \centering
         \includegraphics[width=\textwidth]{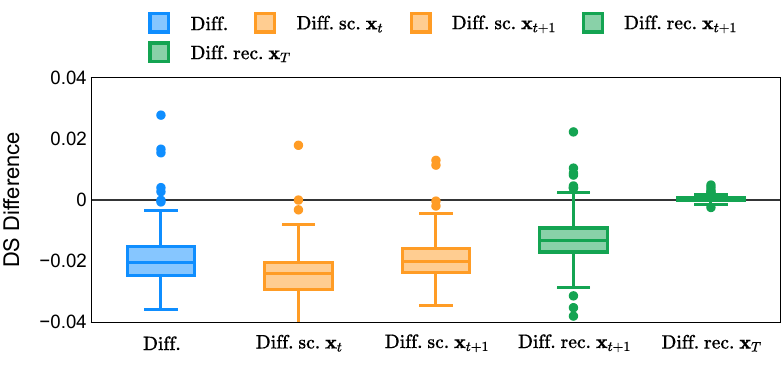}
         \caption{DS difference for prostate MR}
         \label{fig:first_last_step_diff_male_pelvic_mr_DDIM_DS}
     \end{subfigure}
     \begin{subfigure}[b]{0.45\textwidth}
         \centering
         \includegraphics[width=\textwidth]{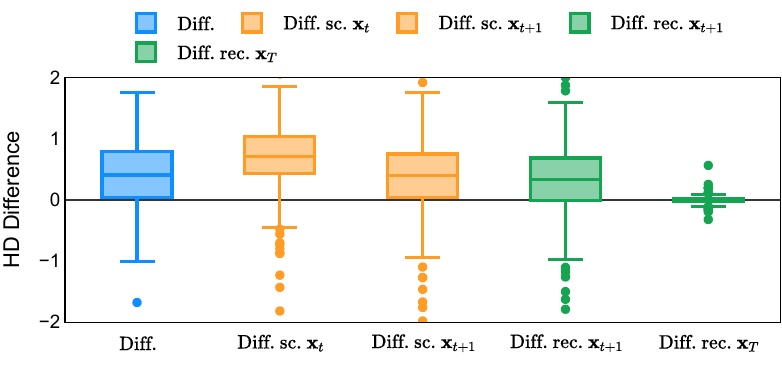}
         \caption{HD difference for prostate MR}
         \label{fig:first_last_step_diff_male_pelvic_mr_DDIM_HD}
     \end{subfigure}
     \begin{subfigure}[b]{0.45\textwidth}
         \centering
         \includegraphics[width=\textwidth]{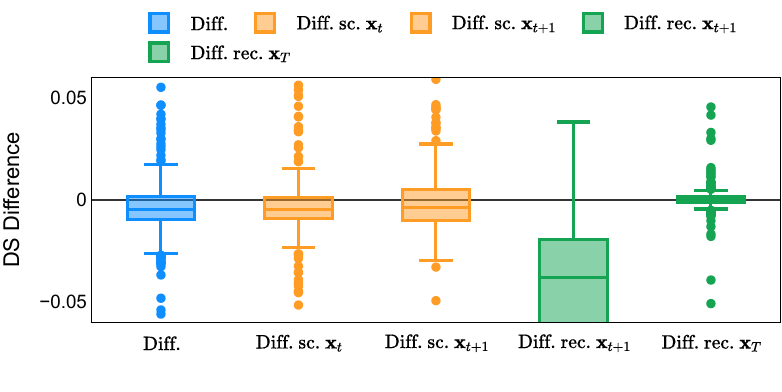}
         \caption{DS difference for brain MR}
         \label{fig:first_last_step_diff_brats2021_mr_DDIM_DS}
     \end{subfigure}
     \begin{subfigure}[b]{0.45\textwidth}
         \centering
         \includegraphics[width=\textwidth]{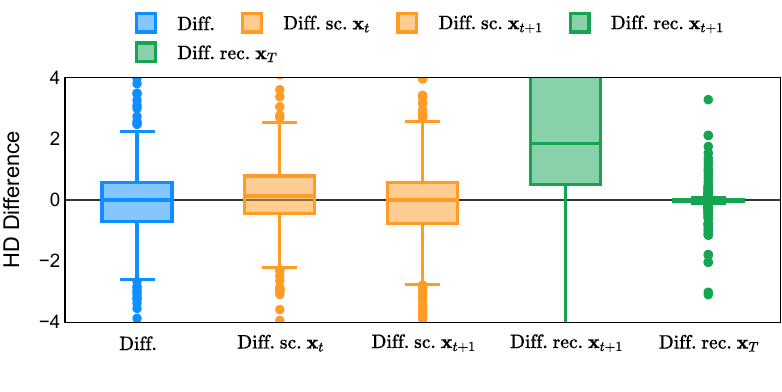}
         \caption{HD difference for brain MR}
         \label{fig:first_last_step_diff_brats2021_mr_DDIM_HD}
     \end{subfigure}
    \caption{\textbf{Segmentation performance difference between the last step and first step using DDIM.} ``Diff.'' represents standard diffusion. ``Diff. sc. $\x_{t}$'' and ``Diff. sc. $\x_{t+1}$'' represents self-conditioning from~\citet{chen2022analog} and~\citet{watson2023novo}, respectively. ``Diff. rec. $\x_{t+1}$'' and ``Diff. rec. $\x_T$'' represents recycling from~\citet{fu2023importance} and the proposed recycling in this work, respectively. The sampler is DDIM. DS and HD represents Dice score and Hausdorff distance, respectively. The difference is the value at the last step subtracted by the one at the first step. A positive value for Dice score difference or a negative value for Hausdorff distance means improvement.}
    \label{fig:first_last_step_diff_ddim}
\end{figure}
\begin{figure}[!ht]
     \centering
     \begin{subfigure}[b]{0.45\textwidth}
         \centering
         \includegraphics[width=\textwidth]{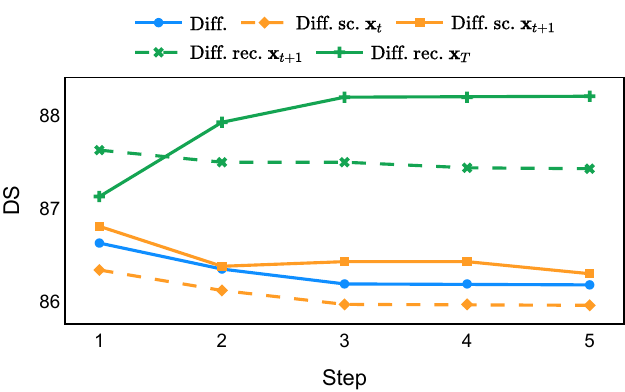}
         \caption{Dice Score for muscle ultrasound}
         \label{fig:per_step_muscle_us_DDIM_DS}
     \end{subfigure}
     \begin{subfigure}[b]{0.45\textwidth}
         \centering
         \includegraphics[width=\textwidth]{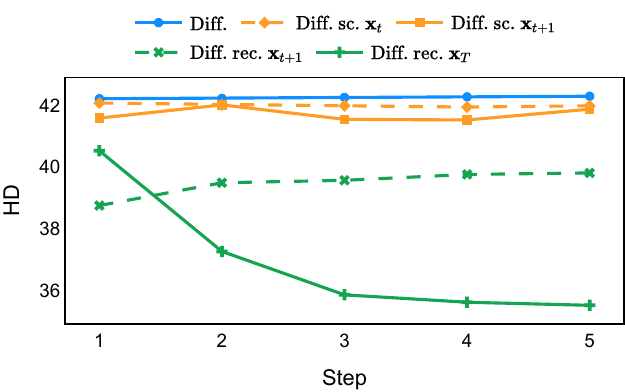}
         \caption{Hausdorff distance for muscle ultrasound}
         \label{fig:per_step_muscle_us_DDIM_HD}
     \end{subfigure}
     \begin{subfigure}[b]{0.45\textwidth}
         \centering
         \includegraphics[width=\textwidth]{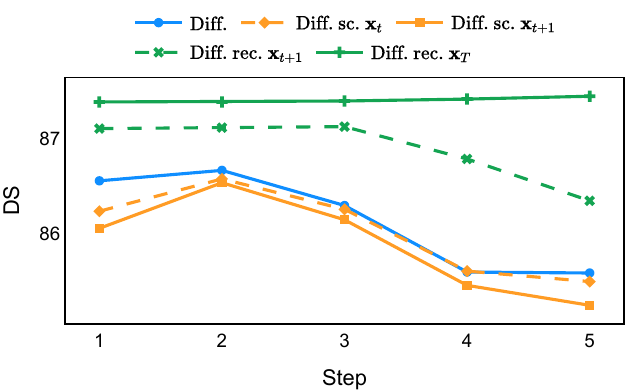}
         \caption{Dice score for abdominal CT}
         \label{fig:per_step_amos_ct_DDIM_DS}
     \end{subfigure}
     \begin{subfigure}[b]{0.45\textwidth}
         \centering
         \includegraphics[width=\textwidth]{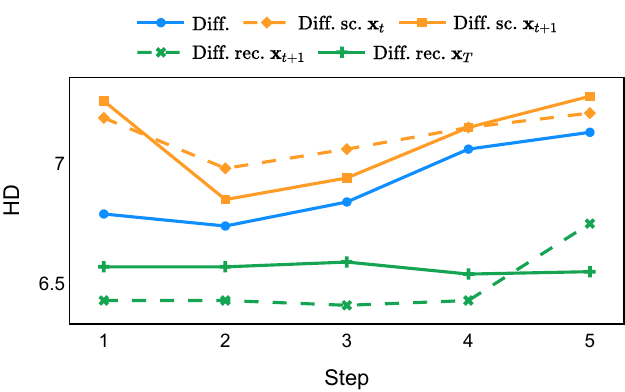}
         \caption{Hausdorff distance for abdominal CT}
         \label{fig:per_step_amos_ct_DDIM_HD}
     \end{subfigure}
     \begin{subfigure}[b]{0.45\textwidth}
         \centering
         \includegraphics[width=\textwidth]{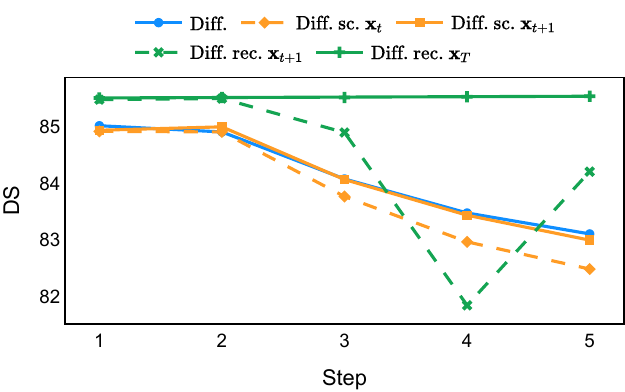}
         \caption{Dice score for prostate MR}
         \label{fig:per_step_male_pelvic_mr_DDIM_DS}
     \end{subfigure}
     \begin{subfigure}[b]{0.45\textwidth}
         \centering
         \includegraphics[width=\textwidth]{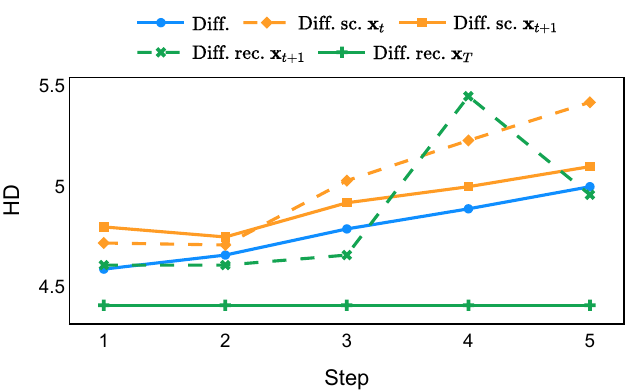}
         \caption{Hausdorff distance for prostate MR}
         \label{fig:per_step_male_pelvic_mr_DDIM_HD}
     \end{subfigure}
     \begin{subfigure}[b]{0.45\textwidth}
         \centering
         \includegraphics[width=\textwidth]{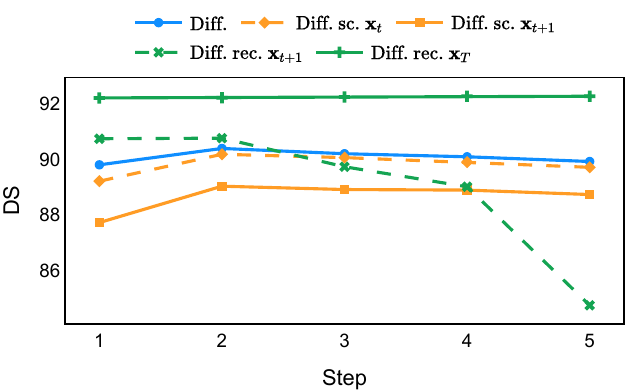}
         \caption{Dice score for brain MR}
         \label{fig:per_step_brats2021_mr_DDIM_DS}
     \end{subfigure}
     \begin{subfigure}[b]{0.45\textwidth}
         \centering
         \includegraphics[width=\textwidth]{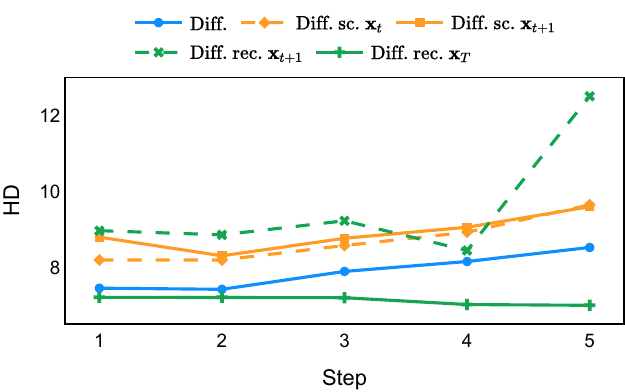}
         \caption{Hausdorff distance for brain MR}
         \label{fig:per_step_brats2021_mr_DDIM_HD}
     \end{subfigure}
    \caption{\textbf{Segmentation performance per step.} ``Diff.'' represents standard diffusion. ``Diff. sc. $\x_{t}$'' and ``Diff. sc. $\x_{t+1}$'' represents self-conditioning from~\citet{chen2022analog} and~\citet{watson2023novo}, respectively. ``Diff. rec. $\x_{t+1}$'' and ``Diff. rec. $\x_T$'' represents recycling from~\citet{fu2023importance} and the proposed recycling in this work, respectively. The sampler is DDIM.}
    \label{fig:per_step_ddim}
\end{figure}
\afterpage{
\subsection{Comparison to Non-diffusion Models}
\begin{figure}[!ht]
     \centering
     \includegraphics[width=0.7\textwidth]{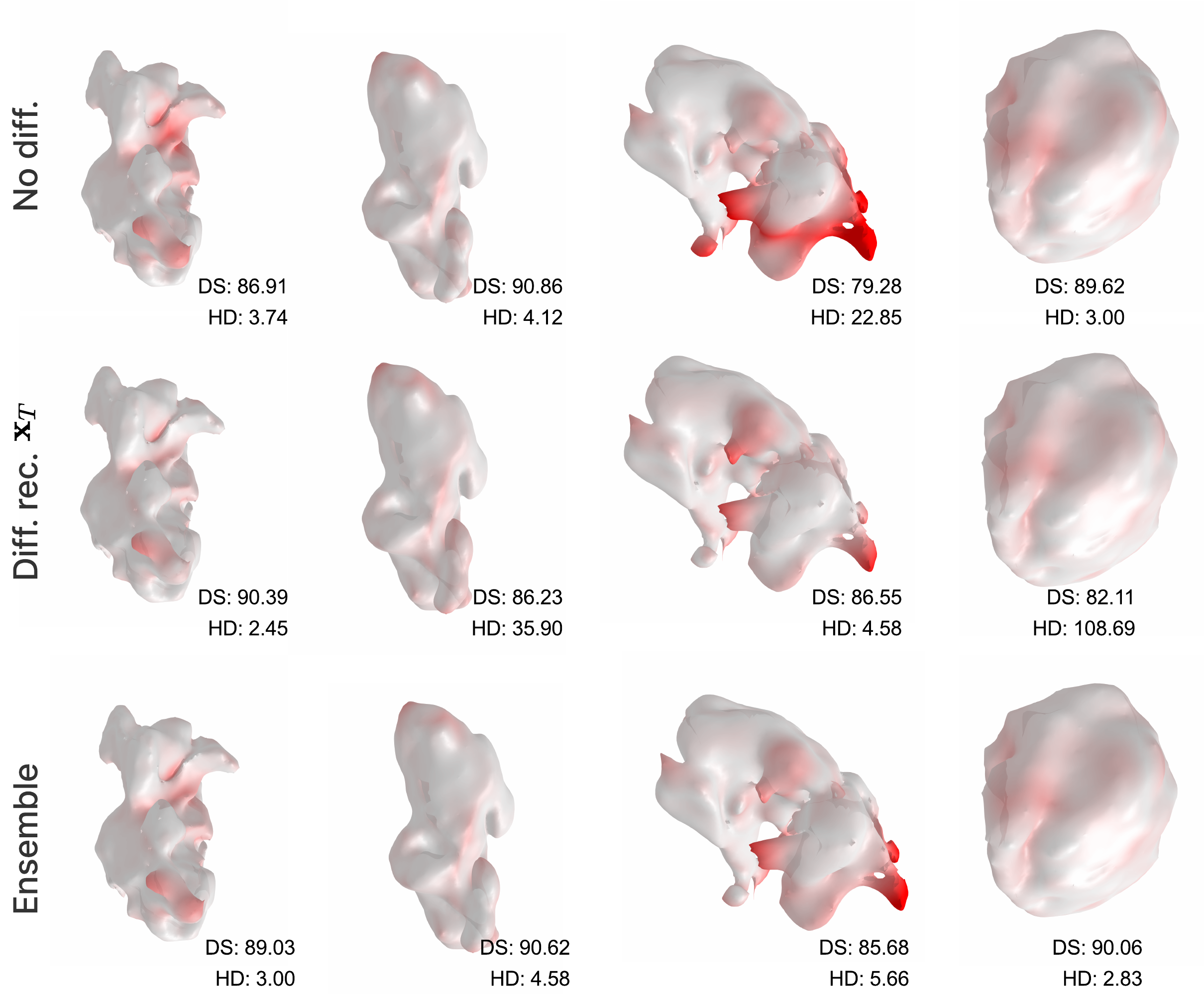}
    \caption{\textbf{Segmentation error of non-diffusion-based and diffusion-based models for tumour in brain MR.} ``No diff.'' represents non-diffusion model. ``Diff. rec. $\x_T$'' represents the diffusion model with proposed recycling. The ground truth segmentation is visualised. For each point on the surface, the distance to the surface of predicted segmentation is calculated and displayed with red color. The Dice score (DS) and Hausdorff distance (HD) for each sample are labeled at bottom.}
    \label{fig:brain_mr_3d}
\end{figure}
\clearpage
}
\afterpage{
\subsection{Ablation Studies}
\begin{table}[!ht]
\centering
\caption{\textbf{Diffusion with different training variance schedule.} ``Diff.'' represents standard diffusion. ``Diff. sc. $\x_{t}$'' and ``Diff. sc. $\x_{t+1}$'' represents self-conditioning from~\citet{chen2022analog} and~\citet{watson2023novo}, respectively. ``Diff. rec. $\x_{t+1}$'' and ``Diff. rec. $\x_T$'' represents recycling from~\citet{fu2023importance} and the proposed recycling in this work, respectively. ``T'' represents the length of variance schedule during training. The best results are in bold and a underline indicates the difference to the second best is significant with p-value $<0.05$.}
\label{tab:timesteps_ablation}

\begin{subtable}[h]{\textwidth}
\centering
\begin{tabular}{c|l|cc|cc}
\toprule
\multirow{2}{*}{T} & \multirow{2}{*}{Method} & \multicolumn{2}{c|}{DDPM} & \multicolumn{2}{c}{DDIM} \\ \cline{3-6}
& & DS $\uparrow$ & HD $\downarrow$ & DS $\uparrow$ & HD $\downarrow$ \\
\midrule
\multirow{5}{*}{1001} & Diff. & 86.60 $\pm$ 12.38 & 41.11 $\pm$ 35.48 & 86.18 $\pm$ 12.41 & 42.31 $\pm$ 35.82 \\
 & Diff. sc. $\x_{t}$ & 86.35 $\pm$ 14.14 & 40.42 $\pm$ 37.53 & 85.96 $\pm$ 13.78 & 42.00 $\pm$ 36.76 \\
 & Diff. sc. $\x_{t+1}$ & 87.14 $\pm$ 11.48 & 39.24 $\pm$ 32.83 & 86.30 $\pm$ 11.49 & 41.89 $\pm$ 32.72 \\
 & Diff. rec. $\x_{t+1}$ & 87.44 $\pm$ 12.39 & 39.68 $\pm$ 36.21 & 87.43 $\pm$ 12.25 & 39.82 $\pm$ 35.39 \\
 & Diff. rec. $\x_{T}$ & 88.23 $\pm$ 11.69 & 35.37 $\pm$ 31.79 & 88.21 $\pm$ 11.70 & 35.52 $\pm$ 31.91 \\
\hline
\multirow{5}{*}{5}  & Diff. & 87.81 $\pm$ 10.98 & 37.39 $\pm$ 31.17 & 87.76 $\pm$ 11.00 & 37.56 $\pm$ 31.34 \\
 & Diff. sc. $\x_{t}$ & 88.11 $\pm$ 11.06 & 35.94 $\pm$ 30.13 & 88.20 $\pm$ 10.73 & 35.57 $\pm$ 29.68 \\
 & Diff. sc. $\x_{t+1}$ & 87.61 $\pm$ 10.88 & 37.76 $\pm$ 29.91 & 88.09 $\pm$ 10.66 & 35.73 $\pm$ 29.26 \\
 & Diff. rec. $\x_{t+1}$ & 88.19 $\pm$ 10.60 & 36.10 $\pm$ 30.38 & 87.83 $\pm$ 11.01 & 37.22 $\pm$ 30.55 \\
 & Diff. rec. $\x_{T}$ & \underline{\textbf{89.01 $\pm$ 10.79}} & \underline{\textbf{33.70 $\pm$ 30.29}} & \underline{\textbf{88.80 $\pm$ 11.54}} & \textbf{34.26 $\pm$ 31.88} \\
\bottomrule
\end{tabular}
\caption{Muscle Ultrasound}
\label{tab:timesteps_ablation_muscle}
\end{subtable}

\hfill

\begin{subtable}[h]{\textwidth}
\centering
\begin{tabular}{c|l|cc|cc}
\toprule
\multirow{2}{*}{T} & \multirow{2}{*}{Method} & \multicolumn{2}{c|}{DDPM} & \multicolumn{2}{c}{DDIM} \\ \cline{3-6}
& & DS $\uparrow$ & HD $\downarrow$ & DS $\uparrow$ & HD $\downarrow$ \\
\midrule
\multirow{5}{*}{1001} & Diff. & 85.25 $\pm$ 5.36 & 7.12 $\pm$ 3.83 & 85.59 $\pm$ 5.24 & 7.13 $\pm$ 3.98 \\
 & Diff. sc. $\x_{t}$ & 86.04 $\pm$ 5.12 & 7.06 $\pm$ 4.20 & 85.50 $\pm$ 5.14 & 7.21 $\pm$ 4.16 \\
 & Diff. sc. $\x_{t+1}$ & 85.86 $\pm$ 5.27 & 6.98 $\pm$ 3.54 & 85.25 $\pm$ 5.42 & 7.28 $\pm$ 3.72 \\
 & Diff. rec. $\x_{t+1}$ & 86.48 $\pm$ 5.24 & 6.69 $\pm$ 4.59 & 86.35 $\pm$ 5.31 & 6.75 $\pm$ 4.55 \\
 & Diff. rec. $\x_{T}$ & \textbf{87.45 $\pm$ 5.43} & \textbf{6.56 $\pm$ 5.44} & \textbf{87.45 $\pm$ 5.43} & \textbf{6.55 $\pm$ 5.43} \\
\hline
\multirow{5}{*}{5} & Diff. & 86.42 $\pm$ 5.00 & 7.09 $\pm$ 4.40 & 86.52 $\pm$ 5.18 & 6.65 $\pm$ 3.88 \\
 & Diff. sc. $\x_{t}$ & 86.68 $\pm$ 4.96 & 7.06 $\pm$ 6.98 & 86.39 $\pm$ 4.87 & 7.12 $\pm$ 6.95 \\
 & Diff. sc. $\x_{t+1}$ & 86.34 $\pm$ 5.33 & 6.69 $\pm$ 3.46 & 86.13 $\pm$ 5.27 & 6.74 $\pm$ 3.55 \\
 & Diff. rec. $\x_{t+1}$ & 87.27 $\pm$ 5.20 & 6.64 $\pm$ 4.69 & 87.27 $\pm$ 5.20 & 6.63 $\pm$ 4.69 \\
 & Diff. rec. $\x_{T}$ & 87.38 $\pm$ 5.46 & 6.71 $\pm$ 4.46 & 87.37 $\pm$ 5.45 & 6.74 $\pm$ 4.49 \\
\bottomrule
\end{tabular}
\caption{Abdominal CT}
\label{tab:timesteps_ablation_abdominal}
\end{subtable}

\end{table}

}
\end{document}